\documentclass[journal]{IEEEtran}
\usepackage{bm}
\usepackage{amsfonts}
\usepackage{amsmath}

\usepackage{graphicx}  
\usepackage{multirow}
\usepackage{booktabs}

\begin{document}

\title{Early Anomaly Detection in Power Systems Based on Random Matrix Theory}

\author{ Xin Shi$^1$, Robert Qiu$^{1,2}$,~\IEEEmembership{Fellow,~IEEE}
\thanks{This work was partly supported by National Key R \& D Program of No. 2018YFF0214705, NSF of China No. 61571296 and (US) NSF Grant No. CNS-1619250.

$^1$ Department of Electrical Engineering,Center for Big Data and Artificial Intelligence, State Energy Smart Grid Research and Development Center, Shanghai Jiaotong University, Shanghai 200240, China.(e-mail: dugushixin@sjtu.edu.cn; rcqiu@sjtu.edu.cn.)

$^2$ Department of Electrical and Computer Engineering,Tennessee Technological University, Cookeville, TN 38505, USA. (e-mail:rqiu@tntech.edu)

}
}

\maketitle

\begin{abstract}
It is important for detecting the anomaly in power systems before it expands and causes serious faults such as power failures or system blackout. With the deployments of phasor measurement units (PMUs), massive amounts of synchrophasor measurements are collected, which makes it possible for the real-time situation awareness of the entire system.
In this paper, based on random matrix theory (RMT), a data-driven approach is proposed for anomaly detection in power systems. First, spatio-temporal data set is formulated by arranging high-dimensional synchrophasor measurements in chronological order. Based on the Ring Law in RMT for the empirical spectral analysis of `signal+noise' matrix, the mean spectral radius (MSR) is introduced to indicate the system states from the macroscopic perspective. In order to realize anomaly declare automatically, an anomaly indicator based on the MSR is designed and the corresponding confidence level $1-\alpha$ is calculated. The proposed approach is capable of detecting the anomaly in an early phase and robust against random fluctuations and measuring errors. Cases on the synthetic data generated from IEEE 300-bus, 118-bus and 57-bus test systems validate the effectiveness and advantages of the approach.
\end{abstract}
\begin{IEEEkeywords}
early anomaly detection, power systems, phasor measurement units (PMUs), random matrix theory (RMT), `signal+noise' matrix
\end{IEEEkeywords}

\IEEEpeerreviewmaketitle

\section{Introduction}
\label{section: Introduction}

\IEEEPARstart{A}{nomaly} in power systems is common, which can easily expand and cause power failures or system blackout if can't be detected in time. The most famous example is the blackout in North East America on the 14th August 2003. In recent years, there have been increasing deployments of phasor measurement units (PMUs), which constitute wide area measurement system (WAMS) \cite{zima2005design} for power systems. Compared with traditional supervisory control and data acquisition (SCADA) system, WAMS can provide synchrophasor measurements with higher sampling rates, which makes it possible for detecting eventualities in a timely manner. For example, the root mean square data of voltage and current can be sampled at the rate of $30$Hz or $60$Hz in the WAMS, while they are obtained only at the rate of $4$Hz or $5$Hz in the SCADA system \cite{ghiocel2014phasor}.

The high sampling rate synchrophasor measurements contain rich information on the operation state of the system, which stimulates the research of advanced data analytics. For example, in \cite{Dasgupta2013Real}, the voltage phasors are used to compute the Lyapunov components to estimate the short-term voltage stability. In \cite{xie2014dimensionality}, the dimensionality of the phasor-measurement-unit (PMU) data is analyzed, and an PCA-based data dimensionality reduction algorithm is proposed for early event detection. In \cite{khan2015parallel}, a parallel detrended fluctuation analysis approach is proposed for fast event detection on massive volumes of PMU data. In \cite{wu2017online}, a density-based detection algorithm is proposed to detect local outliers, which can differentiate high-quality synchrophasor data from the low-quality one during system physical disturbance. In \cite{kim2017wavelet}, a wavelet-based algorithm is proposed for detecting the occurrence of significant changes in time-synchronized voltage and frequency measurements.

Due to the significant deployments of PMUs in WAMS, high-dimensional synchrophasor measurements with fast sampling rate are collected, and the demand for theories capable of processing high-dimensional data has grown dramatically. Random matrix theory (RMT), introduced by Wishart in 1928 \cite{wishart1928generalised}, is an important mathematical tool for statistical analysis of high-dimensional data. As for high-dimensional random matrices, the importance of RMT for statistics comes from the fact that it may be used to correct traditional tests or estimators which fail in the `large $p$, large $n$' setting, where $p$ is the number of parameters (dimensions) and $n$ is the sample size. RMT starts with asymptotic theorems on the distribution of eigenvalues or singular values of random matrices with certain assumptions, and eventually gives macroscopic quantity to indicate the data behavior. The theorems ensure convergence of the empirical eigenvalue distributions to deterministic functions as the matrices grow large, which makes RMT naturally suitable for high-dimensional data analysis. Nowadays, RMT has been widely used in wireless communication \cite{qiu2012cognitive}, finance \cite{saad2013random}, quantum information \cite{chaitanya2015random}, etc. In recent years, some work that makes substantial use of results in the RMT has emerged in the power field. For example, in \cite{he2017big}, an architecture with the application of RMT into smart grid is proposed. In \cite{xu2017correlation}, based on RMT, a data-driven approach to reveal the correlations between various factors and the power system status is proposed. In \cite{Liuwei2016} and \cite{Wuxi2016}, RMT is used for power system transient analysis and steady-state analysis, respectively.

In this paper, based on RMT, a data-driven approach is proposed for anomaly detection in power systems. It tracks the data behavior by moving a window at continuously sampling times. For each moving data window, the empirical eigenvalue distribution (EED) in the complex plane is analyzed and compared with the theoretical limits, and the mean spectral radius (MSR) of the eigenvalues is calculated to indicate the system state in macroscopic. In order to realize anomaly declare automatically in real-time analysis, an anomaly indicator based on the MSR is designed and the corresponding confidence level $1-\alpha$ for each sampling time is calculated. The main contribution of this paper can be summarized as follows:
1) The approach is purely data-driven and does not require making assumptions or simplifications on the complex power systems.
2) The approach is sensitive to the variation of the data behavior, which makes it possible for detecting the anomaly in an early phase.
3) It is theoretically and experimentally justified that the approach is robust against random fluctuations and measuring errors of the data.
4) In the approach, the product of multiplicative data matrices can reinforce the anomaly signals, which makes it much easier for anomaly detection.
5) The proposed approach has fast computing speed, which ensures its application for real-time online analysis.

The rest of this paper is organized as follows. Section \ref{section: analysis} conducts anomaly analysis based on the asymptotic theorem in RMT, and the MSR is introduced as the macroscopic indicator to indicate the system states. In Section \ref{section: theory}, spatio-temporal data set is formulated by arranging high-dimensional synchrophasor measurements in chronological order and specific steps of RMT for anomaly detection are given. The advantages of our approach are systematically analyzed in this section. The synthetic data generated from IEEE 300-bus, 118-bus and 57-bus test systems are used to validate the effectiveness and advantages of our approach in Section \ref{section: case}. Conclusions are presented in Section \ref{section: conclusion}.

\section{Anomaly Analysis Based on Random Matrix Theory}
\label{section: analysis}
In this section, anomaly analysis has been conducted based on RMT. First, asymptotic theorem in RMT is used to analyze the empirical spectral density (ESD) of high-dimensional `signal+noise' matrices, which illustrates the differences of ESD when a system operates in normal and abnormal states. Then the MSR is introduced as the macroscopic indicator to indicate the system states in quantity.

\subsection{Asymptotic Theorem for `Signal+Noise' Matrix}
\label{subsection: statistical properties}
Let ${\bf A}\in \mathbb{C} ^{p \times n}$ be a non-Hermitian random matrix with i.i.d. entries $a_{ij}$. The mean $\mu (a)=0$ and the variance $\sigma^2 (a)=\frac{1}{p}$. The product of $L$ non-Hermitian random matrices can be defined as
\begin{equation}
\label{Eq:matrix_product}
\begin{aligned}
  {\bf Z} = \prod\limits_{k = 1}^L {{{\rm{\bf A}}_{u,k}}}
\end{aligned},
\end{equation}
where ${\bf A}_{u}$ is the singular value equivalent \cite{ipsen2014weak} of ${\bf A}$. As $p,n \to\infty$ but $c=\frac{p}{n}\in (0,1]$, according to the Ring law \cite{guionnet2009single}, the ESD of ${\bf Z}$ converges to the limit with probability density function (PDF)
\begin{equation}
\label{Eq:ring_law}
\begin{aligned}
  {f_{RL}}(a) = \left\{ \begin{array}{l}
\frac{1}{{\pi cL}}{\left| a \right|^{\left( {\frac{2}{L} - 2} \right)}}, \quad {\left( {1 - c} \right)^{\frac{L}{2}}} \le \left| a \right| \le 1\\
0, \qquad\qquad\qquad\qquad {\rm{others}}
\end{array} \right.
\end{aligned}.
\end{equation}

\begin{figure}[htb]
\centering
\begin{minipage}{4.1cm}
\centerline{
\includegraphics[width=1.8in]{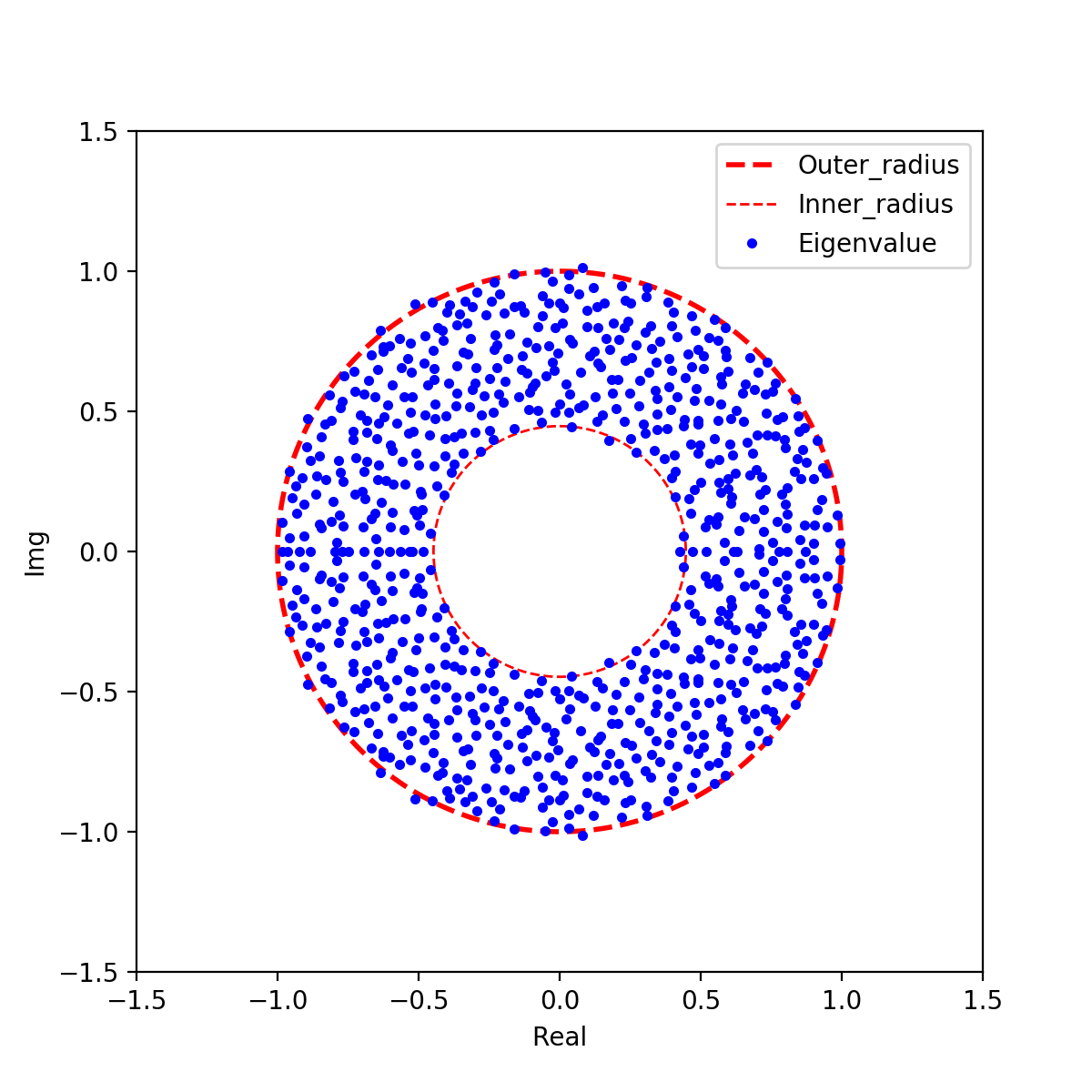}
}
\parbox{5cm}{\small \hspace{1.1cm}(a) Normal state }
\end{minipage}
\hspace{0.2cm}
\begin{minipage}{4.1cm}
\centerline{
\includegraphics[width=1.8in]{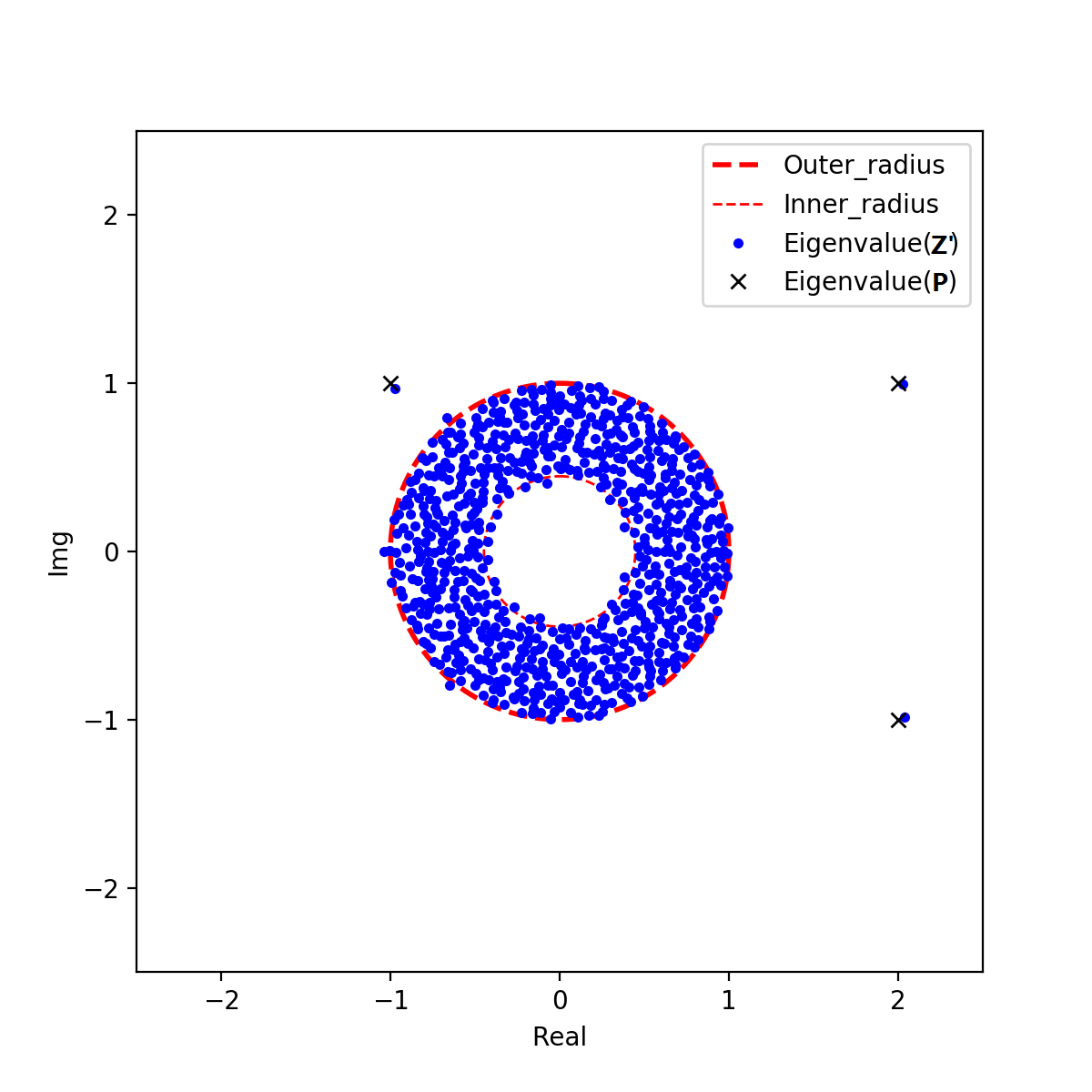}
}
\parbox{5cm}{\small \hspace{1.1cm}(b) Abnormal state }
\end{minipage}
\caption{The ESD of ${{\bf Z'}}$ and its comparison with the theoretical Ring law under both normal and abnormal system states. (a) ${\bf A'}=\frac{1}{\sqrt p}{\bf A}$, where $\bf A$ is a $800\times 1000$ standard random gaussian matrix and $p=800$. (b) ${\bf A'}$ is considered of the type $\frac{1}{\sqrt p}{\bf A}+{\bf P}$, where $\bf A$ is a $800\times 1000$ standard random gaussian matrix, $p=800$, and ${\bf P}=diag(2-i, -1+i, 2+i, 0,\cdots,0)$. The outliers of ${\bf Z'}$ are close to the eigenvalues of $\bf P$, each of which is marked with a cross.}
\label{fig:ringlaw_single}
\end{figure}

1) $L=1$: We first consider the case $L=1$ in the Ring law. Let ${\bf A'}\in {\mathbb{C}}^{p\times n}$ be a high-dimensional data matrix and we apply the Ring law for ${\bf A'}$. In normal state, ${\bf A'}$ is considered as a standard random matrix scaled by $\frac{1}{\sqrt{p}}$ (i.e., ${\bf A'}=\frac{1}{\sqrt{p}}{\bf A}$, where $\bf A$ is a random matrix with the mean $\mu (a)=0$ and the variance $\sigma^2 (a)=1$, and satisfying the singular value equivalent hypothesis), and ${\bf Z'}=\frac{1}{\sqrt p}{{\bf A}_u}$ in equation (\ref{Eq:matrix_product}). Then the ESD of ${{\bf Z'}}$ converges to the limiting spectral density ${f_{RL}}(a')$, which is shown in Figure \ref{fig:ringlaw_single}(a). In the complex plane, the blue dots represent the eigenvalues of ${{\bf Z'}}$, the radius of the inner circle of the ring is $(1-c)^{\frac{1}{2}}$ and the radius of the outer circle of the ring is unity. However, what will happen in abnormal state? Here, ``abnormal'' means signals occur in ${\bf A'}$ and the correlations among the entries $a'_{ij}$ have been changed. Then ${\bf A'}$ is considered of the type $\frac{1}{\sqrt p}{{\bf A}}+{\bf P}$, where ${\bf A}$ is a standard random matrix, which represents random noise or small fluctuations, and $\bf P$ is a low-rank matrix which represents anomaly signals. Figure \ref{fig:ringlaw_single}(b) shows the ESD of ${{\bf Z'}}=\frac{1}{\sqrt p}{{\bf A}_u}+{\bf P}$ does not converge to the Ring law and the extreme eigenvalues (outliers) caused by anomaly signals are out of the outer circle of the ring. More surprising facts are that the outliers are in the neighborhood of the eigenvalues of ${\bf P}$, which has been proved in \cite{benaych2016outliers}.

\begin{figure}[htb]
\centering
\begin{minipage}{4.1cm}
\centerline{
\includegraphics[width=1.8in]{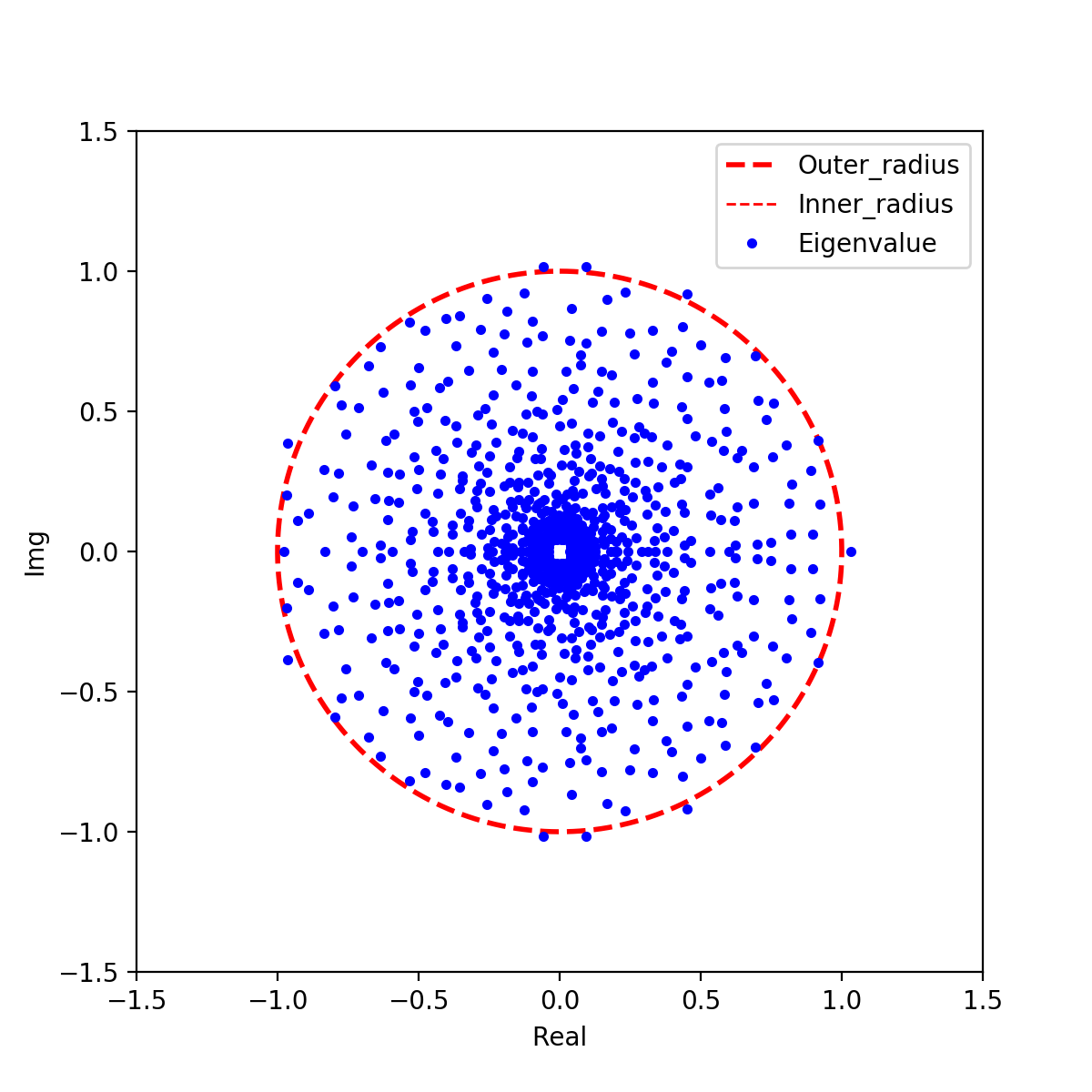}
}
\parbox{5cm}{\small \hspace{1.1cm}(a) Normal state }
\end{minipage}
\hspace{0.2cm}
\begin{minipage}{4.1cm}
\centerline{
\includegraphics[width=1.8in]{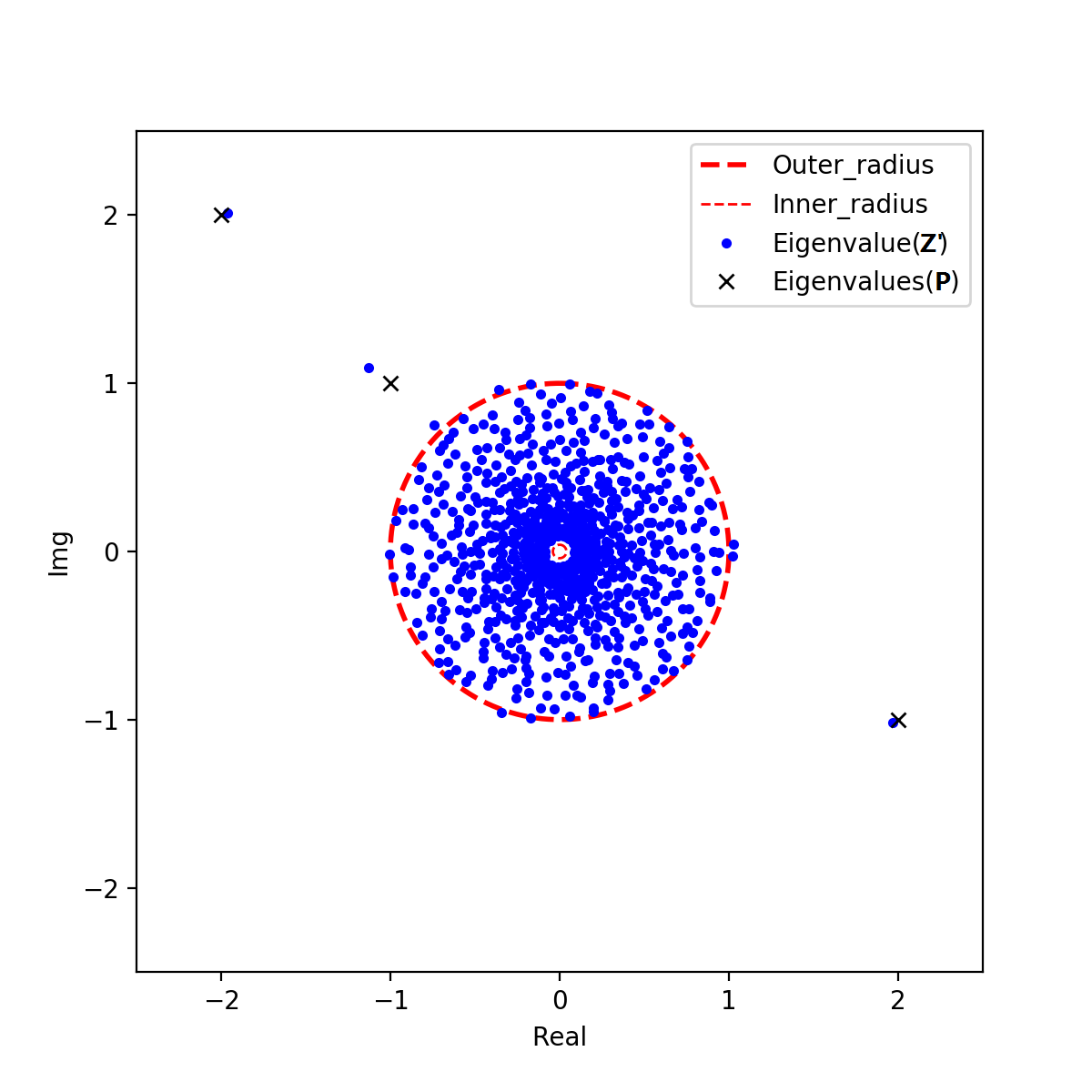}
}
\parbox{5cm}{\small \hspace{1.1cm}(b) Abnormal state }
\end{minipage}
\caption{The ESD of ${{\bf Z'}}$ and its comparison with the theoretical Ring law under both normal and abnormal system states. (a) ${{\bf A'}_k}=\frac{1}{\sqrt p}{{\bf A}_k}\;(k=1,2,3,4)$, where ${\bf A}_k$ is a standard $800\times 1000$ random gaussian matrix and $p=800$. (b) ${{\bf A'}_k}$ is considered of the type ${\frac{1}{\sqrt p}{\bf A}_k}+{{\bf P}_k}\;(k=1,2,3,4)$, where ${{\bf A}_k}$ is a standard $800\times 1000$ random gaussian matrix, $p=800$, ${{\bf P}_{1}}=diag(-1+i, -2, 1, 0,\cdots,0)$, ${{\bf P}_{2}}=diag(1, 1-i, 1, 0,\cdots,0)$, ${{\bf P}_{3}}=diag(1, 1, 2-i, 0,\cdots,0)$, and ${{\bf P}_{4}}=diag(1, 1, 1, 0,\cdots,0)$. The outliers of ${{\bf Z'}}$ are close to the eigenvalues of $\bf P$, each of which is marked with a cross.}
\label{fig:ringlaw_multiple}
\end{figure}

2) $L>1$: Furthermore, we explore the case $L>1$ in the Ring law. Let ${{\bf A'}_k}\in {\mathbb{C}}^{p\times n}\;(k=1,2,\cdots,L)$ be a high-dimensional data matrix and we apply the Ring law for the product of ${{\bf A'}_k}$. In normal state, ${{\bf A'}_k}$ is considered as a standard random matrix scaled by $\frac{1}{\sqrt{p}}$ (i.e., ${{\bf A'}_k}=\frac{1}{\sqrt{p}}{\bf A}_k$, where ${\bf A}_k$ is a random matrix with the mean $\mu (a)=0$ and the variance $\sigma^2 (a)=1$, and satisfying the singular value equivalent hypothesis), and ${{\bf Z'}}=\prod\limits_{k = 1}^L {\frac{1}{\sqrt{p}}{{\rm{\bf A}}_{u,k}}}$ according to equation (\ref{Eq:matrix_product}). Then the ESD of ${{\bf Z'}}$ converges to the limiting spectral density ${f_{RL}}(a')$, which is shown in Figure \ref{fig:ringlaw_multiple}(a). In the complex plane, the blue dots represent the eigenvalues of ${{\bf Z'}}$, the radius of the inner circle of the ring is $(1-c)^{\frac{L}{2}}$ and the radius of the outer circle of the ring is unity.

\begin{figure}[htb]
\centering
\begin{minipage}{4.1cm}
\centerline{
\includegraphics[width=1.8in]{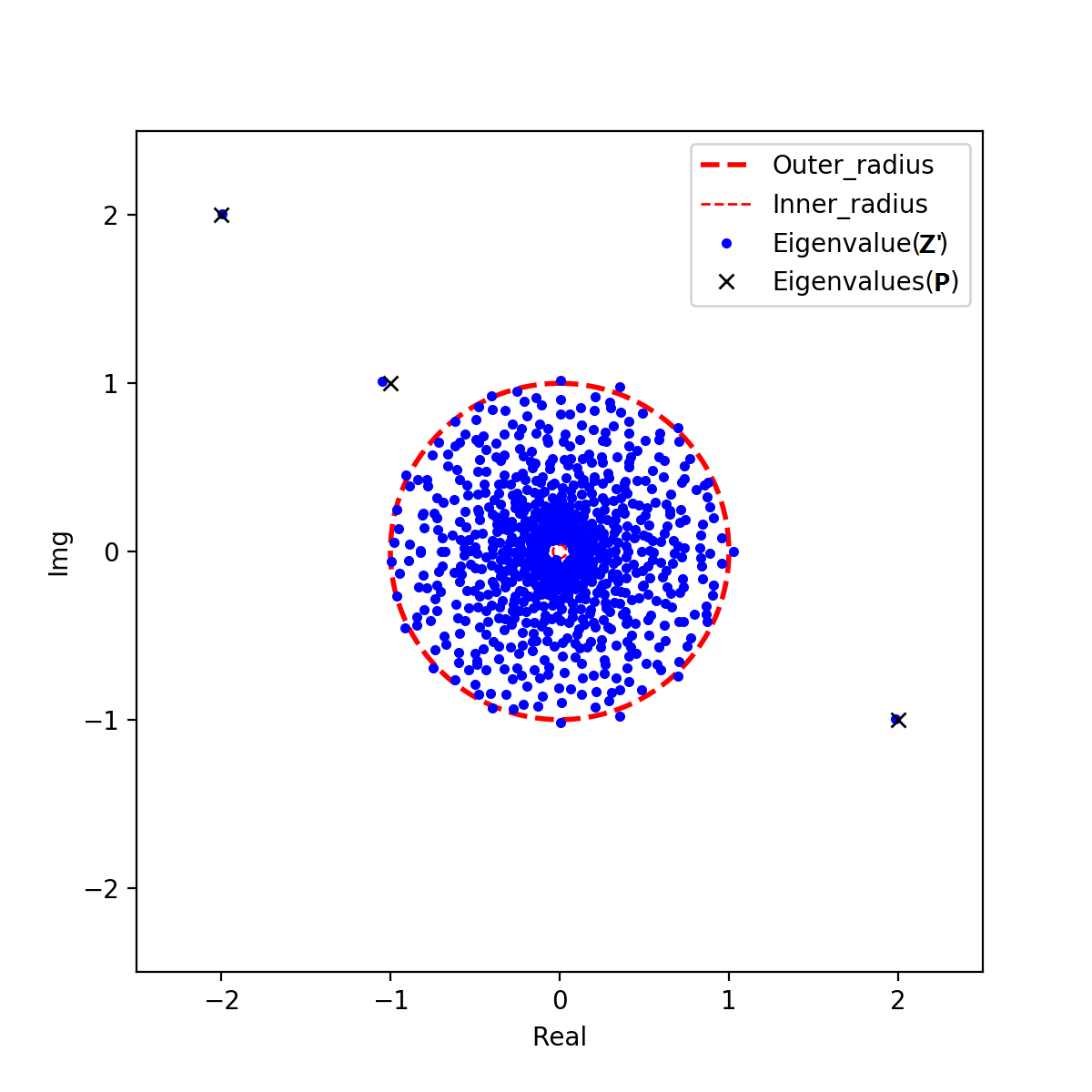}
}
\parbox{5cm}{\small \hspace{2.0cm}(a) }
\end{minipage}
\hspace{0.2cm}
\begin{minipage}{4.1cm}
\centerline{
\includegraphics[width=1.8in]{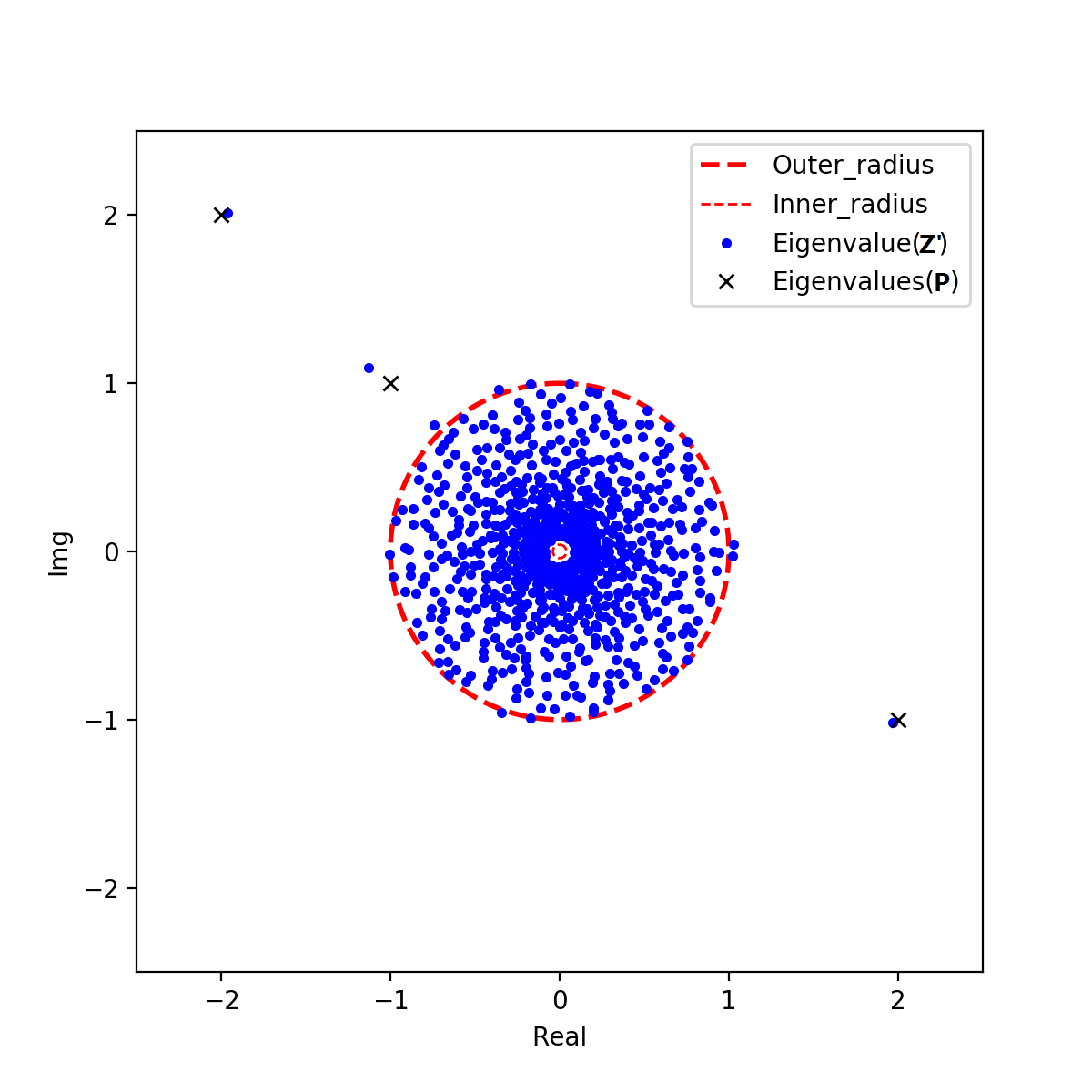}
}
\parbox{5cm}{\small \hspace{2.0cm}(b) }
\end{minipage}
\caption{(a) The ESD of ${\bf Z'}=p^{-2}{{\bf A}_{u,1}}\cdots{{\bf A}_{u,4}}+{{\bf P}_1}\cdots{{\bf P}_4}$, where ${{\bf A}_k}\;(k=1,2,3,4)$ is a standard $800\times 1000$ random gaussian matrix and ${\bf A}_{u,k}$ is the singular value equivalent of ${{\bf A}_k}$, $p=800$, ${{\bf P}_{1}}=diag(-1+i, -2, 1, 0,\cdots,0)$, ${{\bf P}_{2}}=diag(1, 1-i, 1, 0,\cdots,0)$, ${{\bf P}_{3}}=diag(1, 1, 2-i, 0,\cdots,0)$, and ${{\bf P}_{4}}=diag(1, 1, 1, 0,\cdots,0)$. The outliers are close to the eigenvalues of the deterministic term ${{\bf P}_1}\cdots{{\bf P}_4}$. (b) The ESD of ${\bf Z'}=p^{-2}{{\bf A}_{u,1}}\cdots{{\bf A}_{u,4}}+{\bf M}+{{\bf P}_1}\cdots{{\bf P}_4}$, and the `mixed' terms $\bf M$ do not affect the asymptotic location of the outliers.}
\label{fig:ringlaw_multiple_comp}
\end{figure}

In abnormal state, ${{\bf A'}_k}$ is considered of the type $\frac{1}{\sqrt{p}}{{\bf A}_k}+{\bf P}_k$, where ${\bf A}_k$ is a standard random matrix satisfying the singular value equivalent hypothesis, which represents random noise or small fluctuations, and ${\bf P}_k$ is a low-rank matrix which represents anomaly signals. Then the product of ${{\bf A'}_k}$ is calculated as
\begin{equation}
\label{Eq:les}
\begin{aligned}
  {\bf Z'}&=\prod\limits_{k = 1}^L ({\frac{1}{\sqrt{p}}{{\bf A}_{u,k}}}+{\bf P}_k) \\
  &=p^{-L/2}{{\bf A}_{u,1}}\cdots{{\bf A}_{u,L}}+ {\bf M} +{{\bf P}_1}\cdots{{\bf P}_L}
\end{aligned},
\end{equation}
where $p^{-L/2}{{\bf A}_{u,1}}\cdots{{\bf A}_{u,L}}$ is the random term, ${{\bf P}_1}\cdots{{\bf P}_L}$ is the deterministic term, and $\bf M$ represents the `mixed' terms, each containing at least one random factor and one deterministic factor. Figure \ref{fig:ringlaw_multiple}(b) shows the ESD of $\bf Z'$ does not converge to the Ring law and some extreme eigenvalues (outliers) caused by anomaly signals are out of the outer circle of the ring. More surprising facts are that the outliers are only determined by the deterministic term ${{\bf P}_1}\cdots{{\bf P}_L}$ and the `mixed' terms $\bf M$ do not affect the asymptotic location of the outliers, which are shown in Figure \ref{fig:ringlaw_multiple_comp}. Similar conclusions for the Circular law can be found in \cite{coston2017outliers}. Thus, the product of multiple `signal+noise' matrices can reinforce the signals, which makes it much easier to detect the anomaly.
\subsection{Macroscopic Indicator}
\label{subsection: statistical indicator}
\begin{figure}[htb]
\centering
\begin{minipage}{4.1cm}
\centerline{
\includegraphics[width=1.8in]{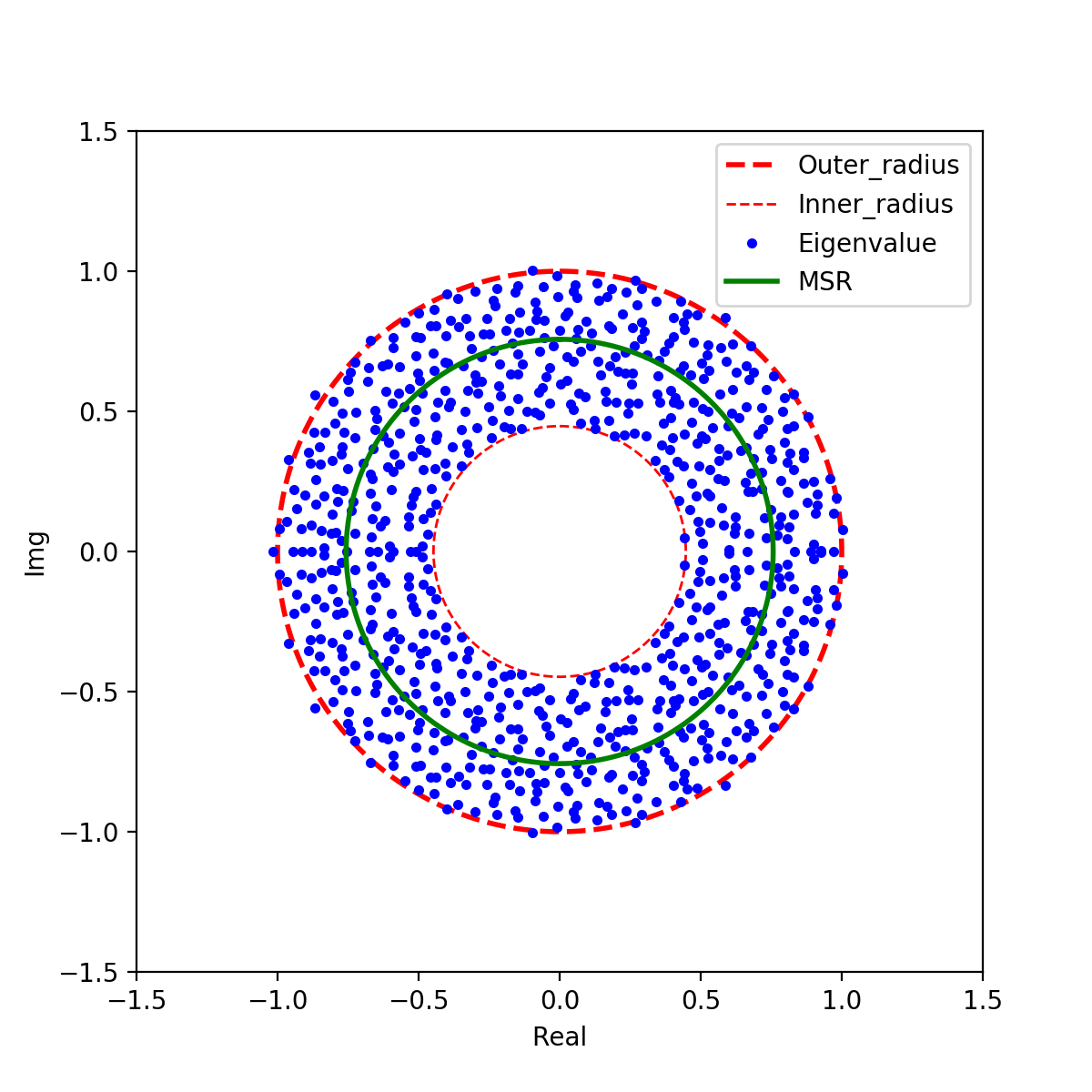}
}
\parbox{5cm}{\small \hspace{1.1cm}(a) Normal state }
\end{minipage}
\hspace{0.2cm}
\begin{minipage}{4.1cm}
\centerline{
\includegraphics[width=1.8in]{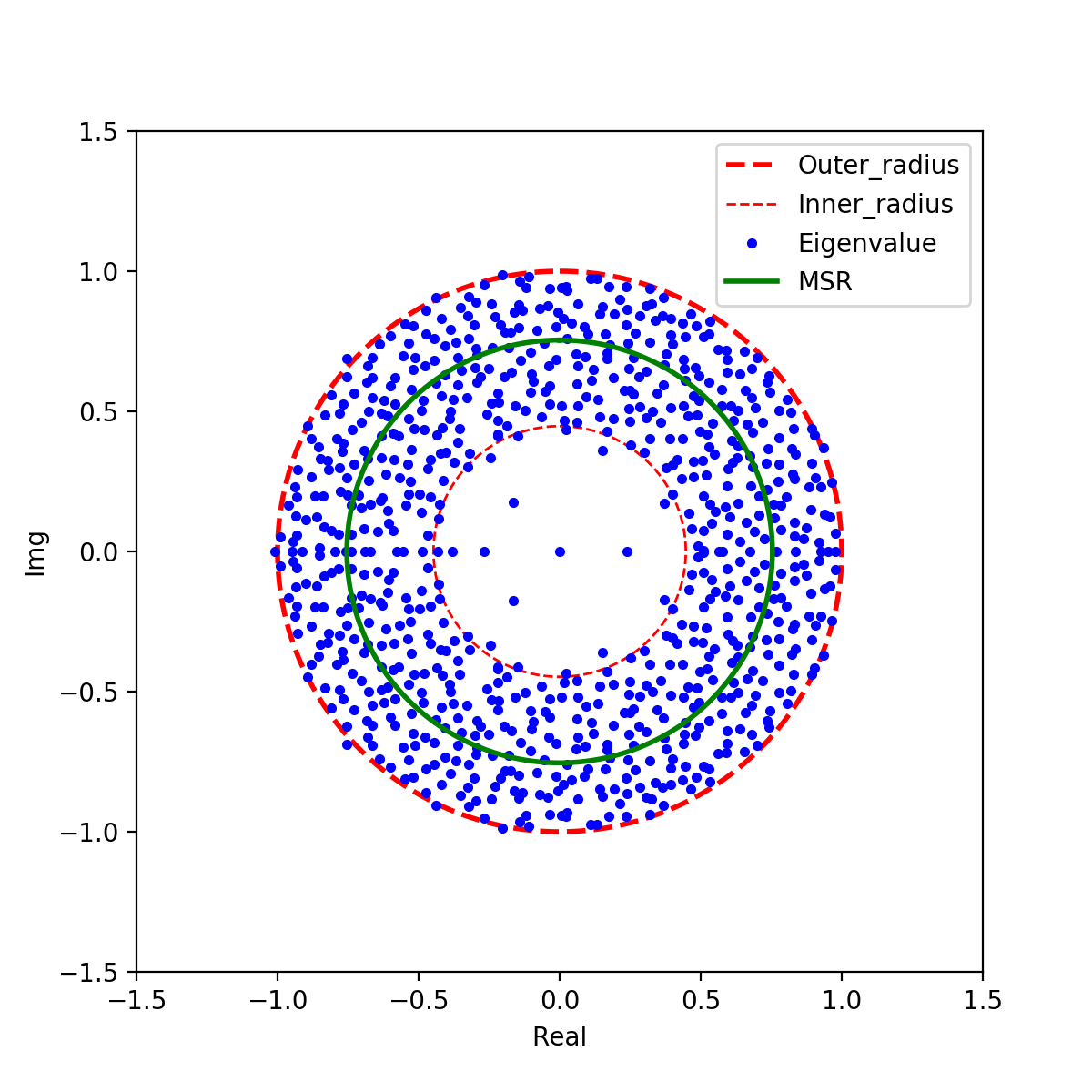}
}
\parbox{5cm}{\small \hspace{1.1cm}(b) Abnormal state }
\end{minipage}
\caption{(a) In normal state, the calculated MSR is $0.7597$. (b) In abnormal state, the value of MSR is $0.7343$.}
\label{fig:ringlaw_single_indicator}
\end{figure}

Based on the analysis in Section \ref{subsection: statistical properties}, it can be concluded that the ESDs are different for a high-dimensional random matrix with or without anomaly signals, which inspires us to investigate the statistics regarding the empirical eigenvalues in the complex plane to indicate the data behavior in macroscopic quantity. The mean spectral radius (MSR) is the mean distribution radius of the eigenvalues of the product $\bf Z$ in equation (\ref{Eq:matrix_product}), which is defined as
\begin{equation}
\label{Eq:msr}
\begin{aligned}
  {\kappa _{\rm {MSR}}} = \frac{1}{p}\sum\limits_{i = 1}^p {\left| {{\lambda _i}} \right|}
\end{aligned},
\end{equation}
where $\lambda _i\;(i=1,\cdots,p)$ are the eigenvalues of $\bf Z$ and  $| \lambda _i |$ is the radius of $\lambda _i$ in the complex plane. It can be used to measure the distribution of the eigenvalues in macroscopic quantity.
For example, the MSR of the eigenvalues of the standard $\bf Z'$ (i.e., $\sigma^2 ({\bf z'}_i)=\frac{1}{p}\;(i=1,\cdots,p)$ for each row of  $\bf Z'$) under both normal and abnormal system states corresponding to $L=1$ in Section \ref{subsection: statistical properties} are shown in Figure \ref{fig:ringlaw_single_indicator}. It can be concluded that the MSR decreases when the system state changes from normal to abnormal.
\section{Random Matrix Theory for Anomaly Detection}
\label{section: theory}
Based on the analysis in Section \ref{section: analysis}, an RMT-based anomaly detection approach is proposed in this section. First, spatio-temporal data set is formulated by arranging high-dimensional synchrophasor measurements in chronological order in power systems. Then, details on the RMT for real-time data analysis are presented, in which a moving data window method is used. Finally, steps of the RMT for anomaly detection are given, and we systematically analyze the advantages of the approach.

\subsection{Spatio-Temporal Data Formulation}
\label{subsection: data formulation}
Assume there are $P$-dimensional measurement variables (such as $P-$dimensional voltage measurements from $P$ PMUs installed in a power system) $(d_1,d_2,...,d_P)\in \mathbb{C}^{1\times P}$. At the sampling time $t_j$, the $P-$dimensional measurements can be formulated as a column vector ${\bf d}(t_j)=(d_1,d_2,...,d_P)^H$. For a series of time $N$, a spatio-temporal data set ${\bf D}\in \mathbb{C}^{P\times N}$ is formulated by arranging these vectors $\bf d$ in chronological order. To be mentioned is that, by stacking the $P$ measurements in a series of time $N$ together, the spatio-temporal data contains the most information on the operating states of the system.

\subsection{Real-Time Data Analysis for Anomaly Detection}
\label{subsection: real_time_analysis}
In real-time analysis, we can move a $p\times n\;(p=P,\;n\le N)$ window on $\bf D$ at continuous sampling times and the last sampling time is the current time, which enables us to track the data behavior in real-time. For example, at the sampling time $t_j$, the obtained data window ${\bf X}(t_j)$ is formulated as
\begin{equation}
\label{Eq:matrix_formulate}
\begin{aligned}
  {\bf{X}}(t_j) = \left( {{\bf{d}}(t_{j - n + 1}),{\bf{d}}(t_{j - n + 2}), \cdots ,{\bf{d}}(t_j)} \right)
\end{aligned},
\end{equation}
where ${\bf d}(t_k)={({d_1,d_2,\cdots,d_p})}^H$ for $t_{j-n+1}\le t_k \le t_j$ is the sampling data at time $t_k$.

For the data matrix ${\bf X}\in\mathbb{C}^{p\times n}$, we convert it into the standard form $\hat{\bf X}$ by
\begin{equation}
\label{Eq:standardize}
\begin{aligned}
  {\hat x_{ij}} = \left( {{x_{ij}} - \mu \left( {{{\bf x}_i}} \right)} \right) \times \frac{{\sigma \left( {{{\hat {\bf x}}_i}} \right)}}{{\sigma \left( {{{\bf x}_i}} \right)}} + \mu \left( {{{\hat {\bf x}}_i}} \right)
\end{aligned},
\end{equation}
where ${\bf x}_i=(x_{i1},x_{i2},...,x_{in})$, $\mu ({\bf\hat x}_i)=0$, and $\sigma ({\bf\hat x}_i)=1$ $(i=1,2,...,p;j=1,2,...,n)$. The singular value equivalent of $\hat{\bf X}$ is introduced as
\begin{equation}
\label{Eq:singular_value_equ}
\begin{aligned}
  {\hat{\bf X}}_{u}=\sqrt{\frac{1}{n}{\hat{\bf X}}{\hat{\bf X}}^{H}}{\bf U}
\end{aligned},
\end{equation}
where ${\bf U}\in\mathbb{C}^{p\times p}$ is a Haar unitary matrix, and ${\hat{\bf X}}_{u}\in\mathbb{C}^{p\times p}$.

According to equation (\ref{Eq:matrix_product}), the product of $L$ (such as $t_{j-L+1}\sim t_j$ for the sampling time) non-Hermitian matrices ${\hat{\bf X}}_{k}$ is obtained by
\begin{equation}
\label{Eq:multiple_product_x}
\begin{aligned}
  {\bf Z}=\prod\limits_{k = 1}^L {\hat{\bf X}}_{u,k}
\end{aligned}.
\end{equation}
Then $\bf Z$ is converted to the standard form $\hat{\bf Z}$ by
\begin{equation}
\label{Eq:standard_z}
\begin{aligned}
  {\hat {\bf z}}_{i}=\frac{{\bf z}_i}{{\sqrt{p}}\sigma ({\bf z}_{i})}
\end{aligned},
\end{equation}
where ${{\bf z}_i}=(z_{i1},\cdots,z_{ip})\;(i=1,\cdots,p)$, $\sigma ({\bf z}_{i})$ is the standard deviation of ${\bf z}_{i}$, and thus $\sigma^{2} ({\hat{\bf z}}_{i})=\frac{1}{p}$. Then the eigenvalues ${\lambda} _{\hat{\bf Z},i}$ of $\hat{\bf Z}$ can be obtained and the MSR $\kappa _{\rm MSR}$ in equation (\ref{Eq:msr}) can be calculated. Since ${\lambda} _{\hat{\bf Z},i}$ is a complicated matrix function of the product $\hat{\bf Z}$, it can be considered as a random variable.

In practice, for a series of time $T\;(T\le N)$, $\kappa _{\rm MSR}$ is generated for each sampling time with continuously moving windows and the last sampling time $t_j$ is considered as the current time. In order to realize anomaly declare automatically in real-time analysis, an anomaly indicator $\eta$ is defined based on the generated $\kappa_{\rm MSR}$. For example, at the sampling time $t_k$, the anomaly indicator $\eta(t_k)$ is calculated as
\begin{equation}
\label{Eq:eta}
\begin{aligned}
  {\eta(t_k)} = {|{\kappa _{\rm MSR}(t_j)}-\kappa _{\rm MSR}({t_{j-1}})|}
\end{aligned},
\end{equation}
where $j-T+1<k\le j$ and $|\cdot|$ is the absolute value function. Here, $\eta$ for the $T$ sampling times is considered to follow a student t-distribution with $T-1$ degree of freedom. Now we convert $\eta$ for the last sampling time $t_j$ into the standard form $\hat\eta$ by
\begin{equation}
\label{Eq:eta_standard}
\begin{aligned}
  {\hat\eta(t_j)} = \frac{|{\eta(t_j)}-\mu (\eta)|}{\sigma (\eta)}
\end{aligned},
\end{equation}
where $\mu (\eta)$ and $\sigma (\eta)$ are the mean and standard deviation of $\eta$ for the $T$ sampling times, and $\hat{\eta}$ follows the standard t-distribution with $T-1$ degree of freedom. We can obtain the confidence level $1-\alpha$ of $\eta(t_j)$ once ${\hat\eta(t_j)}$ and $T-1$ are calculated. For example, let ${\hat\eta(t_j)}=2.364$ and $T-1=100$, then the confidence level $1-\alpha$ for the last sampling time $t_j$ is $98\%$. Thus, an anomaly can be declared automatically by comparing $1-\alpha$ with the threshold ${(1-\alpha)}_{th}$ defined empirically.
\subsection{Anomaly Detection Approach and Its Advantages}
\label{subsection: anomaly_detection_approach}
Based on the research mentioned above, an RMT-based approach is proposed for anomaly detection in power systems. The specific steps are given in Table \ref{Tab: steps_approach}.
\begin{table}[htbp]
\label{Tab: steps_approach}
\centering
\begin{tabular}{p{8.4cm}}   
\toprule[1.0pt]
\textbf {Steps of the RMT-based Anomaly Detection Approach}\\
\hline
1: A spatio-temporal data set ${\bf D}\in{\mathbb{C}^{P\times N}}$ is formulated by arranging $P$ \\
\quad synchrophasor measurements in a series of time $N$ in chronological \\
\quad order.  \\
2: For the sampling time $t_j$: \\
  \quad 2a) Obtain the data matrix ${\bf X}(t_j)$ as in equation (\ref{Eq:matrix_formulate}) by using a $p\times n$ \\
  \quad\quad\; ($p=P,n\le N$) window on $\bf D$; \\
  \quad 2b) Convert ${\bf X}(t_j)$ into the standard form matrix ${\hat{\bf X}}(t_j)$ through \\
  \quad\quad\; equation (\ref{Eq:standardize}); \\
  \quad 2c) The singular value equivalent of ${\hat{\bf X}}(t_j)$ is introduced as ${\hat{\bf X}}_{u}(t_j)$ in  \\
  \quad\quad\; equation (\ref{Eq:singular_value_equ});  \\
  \quad 2d) The product ${\bf Z}(t_j)$ of $L$ non-Hermitian matrices ${\hat{\bf X}}(t_j)$ is calculated \\
  \quad\quad\; through equation (\ref{Eq:multiple_product_x}); \\
  \quad 2e) Convert ${\bf Z}(t_j)$ into the standard form ${\hat{\bf Z}}(t_j)$ through equation (\ref{Eq:standard_z}); \\
  \quad 2f) Calculate the eigenvalues ${\lambda} _{\hat{\bf Z},i}\;(i=1,\cdots,p)$ of $\hat{\bf Z}(t_j)$ and \\
  \quad\quad\; compare the ESD with the theoretical Ring law; \\
  \quad 2g) Calculate $\kappa _{\rm MSR}(t_j)$ through equation (\ref{Eq:msr}) and the corresponding \\
  \quad\quad\; $\eta(t_j)$ in equation (\ref{Eq:eta}); \\
3: Draw the $\kappa _{\rm MSR}-t$ and $\eta-t$ curves for a series of time $N$. \\
4: Calculate the confidence level $1-\alpha$ of $\eta$ for each sampling time and \\
\quad declare anomalies where ${(1-\alpha)}>{(1-\alpha)}_{th}$. \\
\hline
\end{tabular}
\end{table}

Step 1 is conducted for data set formulation. For each sampling time, obtain a moving data window on the data set and the last sampling time is considered as the current time, which is shown in Step 2a. In Step 2b$\sim$2f, the Ring law in RMT is used for the data window analysis. The MSR and corresponding anomaly indicator for each sampling time are calculated in Step 2g. The $\kappa _{\rm MSR}-t$ and $\eta-t$ curves are drawn in Step 3 to indicate the data behavior. In Step 4, the confidence level of $\eta$ for each sampling time is calculated and compared with the threshold to realize anomaly declare automatically.

The RMT-based anomaly detection approach is purely data-driven and suitable for high-dimensional data analysis, such as tens, hundreds or more. It is sensitive to the variation of the data behavior, which makes it possible for detecting the early anomalies. The steps above involve no mechanism models, thus avoiding the errors brought by assumptions and simplifications. For each sampling time, a spatio-temporal data window instead of just the current sampling data is analyzed in the approach. The average result makes it robust against random fluctuations and measuring errors in the data. What's more, our approach is practical for real-time analysis for the fast computing speed.
\section{Case Studies}
\label{section: case}
In this section, the effectiveness of our approach is validated with the synthetic data generated from IEEE 300-bus, 118-bus and 57-bus test systems \cite{5491276}. Detailed information about IEEE 300-bus, 118-bus and 57-bus test systems can be found in case300.m, case118.m and case57.m in Matpower6.0 package \cite{zimmerman2016matpower}. And the simulation environment is MATLAB2016. Three cases in different scenarios were designed: 1) The first case, leveraging the synthetic data from IEEE 300-bus test system, tested the effectiveness of our approach with $L=1$ for anomaly detection, in which an anomaly signal was set by a sudden increase of impedance. 2) In the second case, we tested the effectiveness of our approach with $L>1$ for anomaly detection and compared that with the case $L=1$. In this case, the synthetic data was generated from IEEE 118-bus test system and an increasing anomaly signal was set by gradually increasing the active load. 3) We validated the advantages of our approach in anomaly detection by comparing it with other existing techniques in the last case. In this case, the synthetic data was generated from IEEE 57-bus test system and an increasing anomaly signal was set by gradually increasing the active load.
\subsection{Case Study on $L=1$}
\label{subsection: case_A}
\begin{figure}[!t]
\centerline{
\includegraphics[width=3.0in]{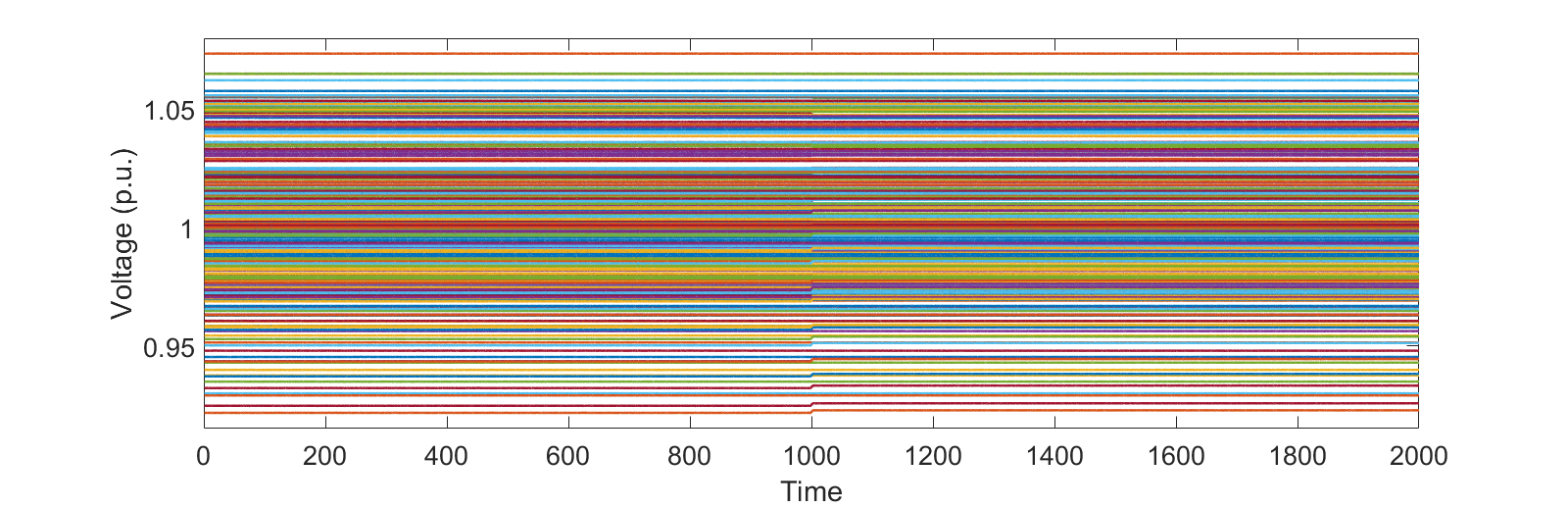}
}
\caption{The synthetic data generated from IEEE 300-bus test system in Case A. An anomaly signal was set at $t_s=1001$.}
\label{fig:case1_data_org}
\end{figure}

In this case, the synthetic data generated from IEEE 300-bus test system contained $300$ voltage measurement variables with sampling $2000$ times. For the generated data set $\bf D$, a little white noise $\bf E$ ($E_{it}\sim N(0,1),\;i=1,\cdots,300;\;t=1,\cdots,2000$) was introduced to represent random fluctuations and measuring errors, i.e., ${\tilde{\bf D}}={\bf D}+\gamma{\bf E}$. The scale of the added white noise is $\gamma=\sqrt{\frac{Tr({\bf D}{\bf D}^H)}{Tr({\bf E}{\bf E}^H)\times{\tau_{SNR}}}}$, where $Tr(\cdot)$ represents the trace function and $\tau_{SNR}$ is the signal-noise-rate. In order to test the effectiveness of our approach with $L=1$ in equation (\ref{Eq:multiple_product_x}), an assumed anomaly signal was set by a sudden increase of impedance from bus $33$ to $34$ and others stayed unchanged, which was shown in Table \ref{Tab: Case1}. The generated data was shown in Figure \ref{fig:case1_data_org}. In the experiments, the size of the moving window was set as $300\times 500$ and the signal-noise-rate $\tau_{SNR}$ was set to be $10000$. The experiments were conducted for $20$ times and the results were averaged.
\begin{table}[!t]
\caption{An Assumed Signal From Bus $33$ to $34$ in Case A.}
\label{Tab: Case1}
\centering
\begin{tabular}{cclc}   
\toprule[1.0pt]
\textbf {fBus} & \textbf {tBus} & \textbf{Sampling Time}& \textbf{Impedance(p.u.)}\\
\hline
\multirow{2}*{33} & \multirow{2}*{34} & $t_s=1\sim 1000$ & 0.2 \\
~ & ~ & $t_s=1001\sim 2000$ & 20 \\
Others & Others & $t_s=1\sim 2000$ & Unchanged \\
\hline
\end{tabular}
\end{table}

\begin{figure}[!t]
\centering
\begin{minipage}{4.1cm}
\centerline{
\includegraphics[width=1.8in]{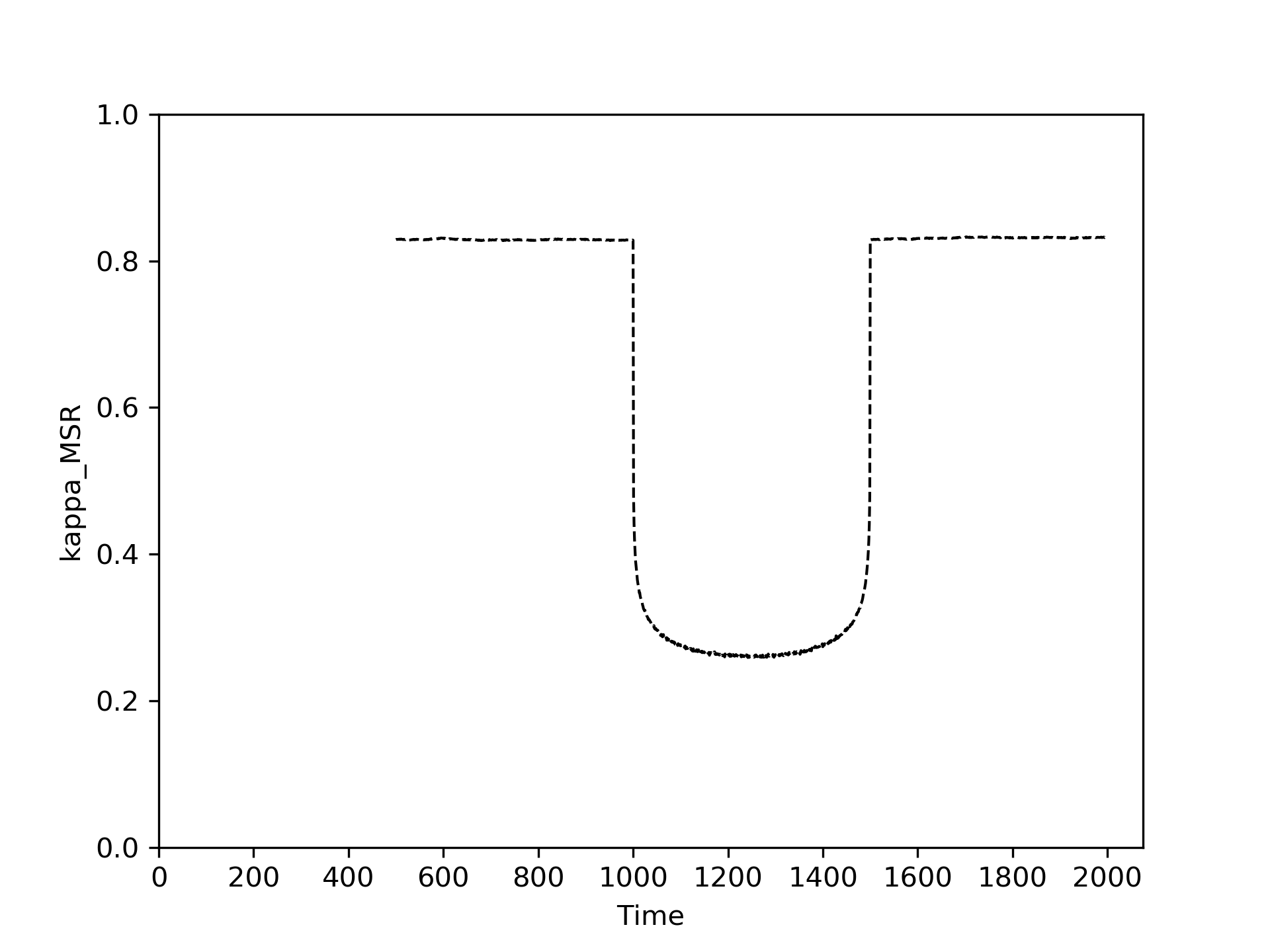}
}
\parbox{5cm}{\small \hspace{1.2cm}(a) $\kappa_{\rm MSR}-t$ curve}
\end{minipage}
\hspace{0.2cm}
\begin{minipage}{4.1cm}
\centerline{
\includegraphics[width=1.8in]{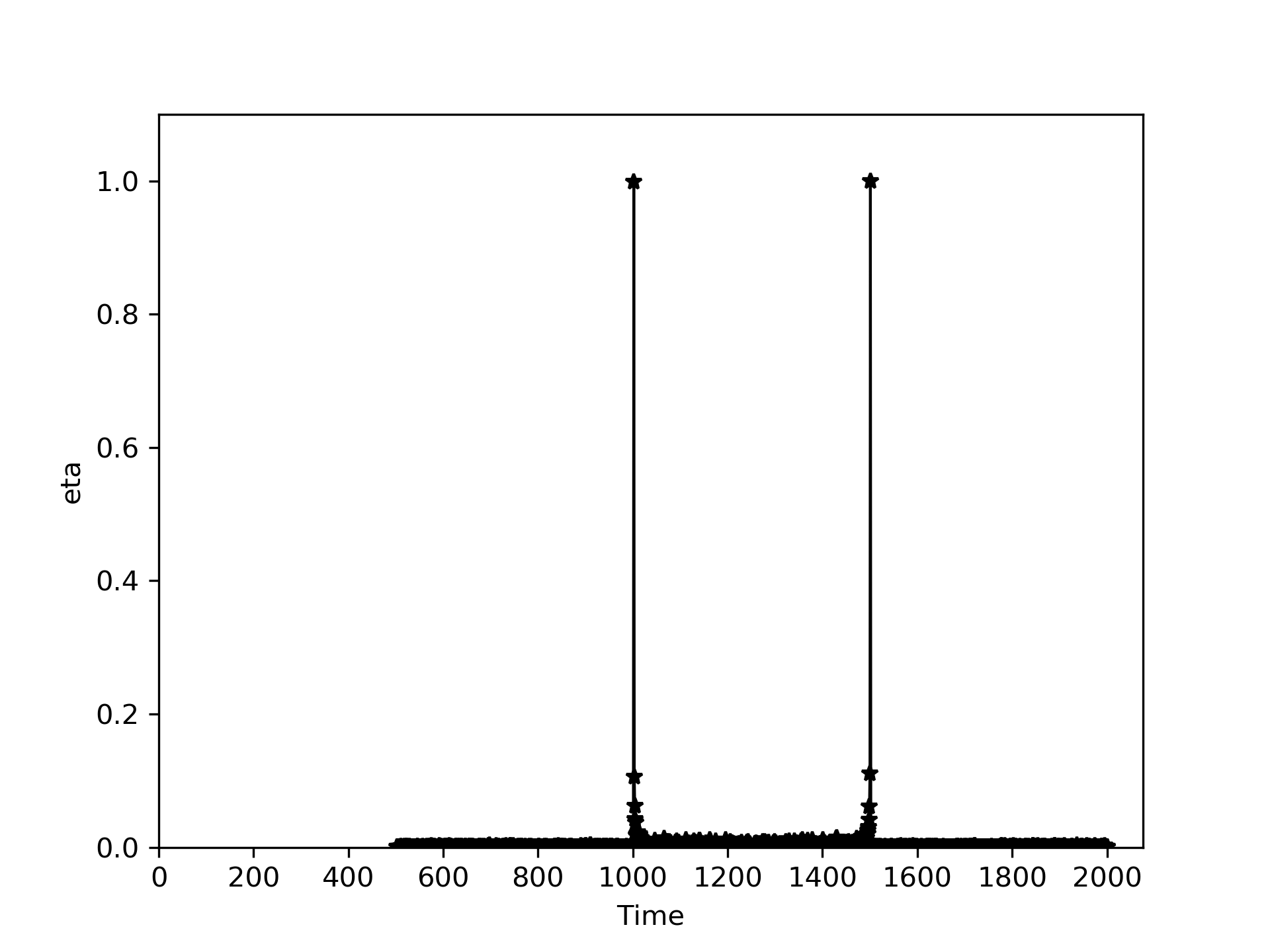}
}
\parbox{5cm}{\small \hspace{1.5cm}(b) $\eta-t$ curve}
\end{minipage}
\caption{The anomaly detection results in Case A.}
\label{fig:case1_indicator}
\end{figure}
The anomaly detection results are shown in Figure \ref{fig:case1_indicator}. The anomaly indicator $\eta$ in Figure \ref{fig:case1_indicator}(b) is normalized into $(0,1]$. It is noted that the $\kappa_{\rm MSR}-t$ curve begins at $t_s=500$, because the initial moving window includes 499 times of historical sampling and the present sampling data. And the $\eta-t$ curve begins at $t_s=501$ for $\eta{(501)}$ is calculated through $\kappa_{\rm MSR}{(500)}$ and $\kappa_{\rm MSR}{(501)}$ in equation (\ref{Eq:eta}). In the real-time calculation of $1-\alpha$ for each data point on the $\eta-t$ curve, a series of $T=100$ sampling times (i.e., $99$ times of historical sampling times and the current sampling time) are considered to follow a student t-distribution. The threshold ${(1-\alpha)}_{th}$ is set to be $98\%$. The anomaly detection processes are shown as follows:

\begin{figure}[!t]
\centering
\begin{minipage}{4.1cm}
\centerline{
\includegraphics[width=1.8in]{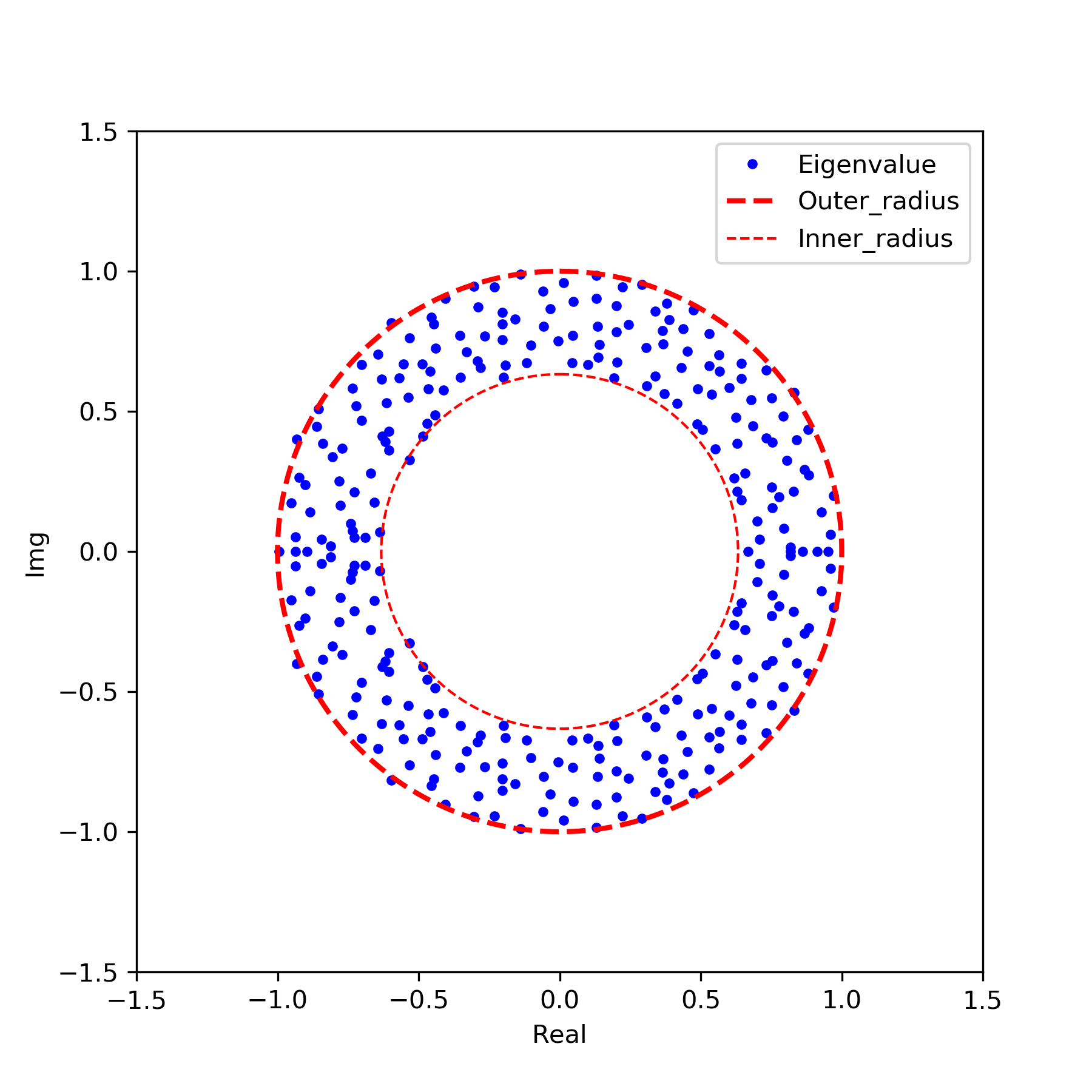}
}
\parbox{5cm}{\small \hspace{1.5cm}(a) $t_s=1000$}
\end{minipage}
\hspace{0.2cm}
\begin{minipage}{4.1cm}
\centerline{
\includegraphics[width=1.8in]{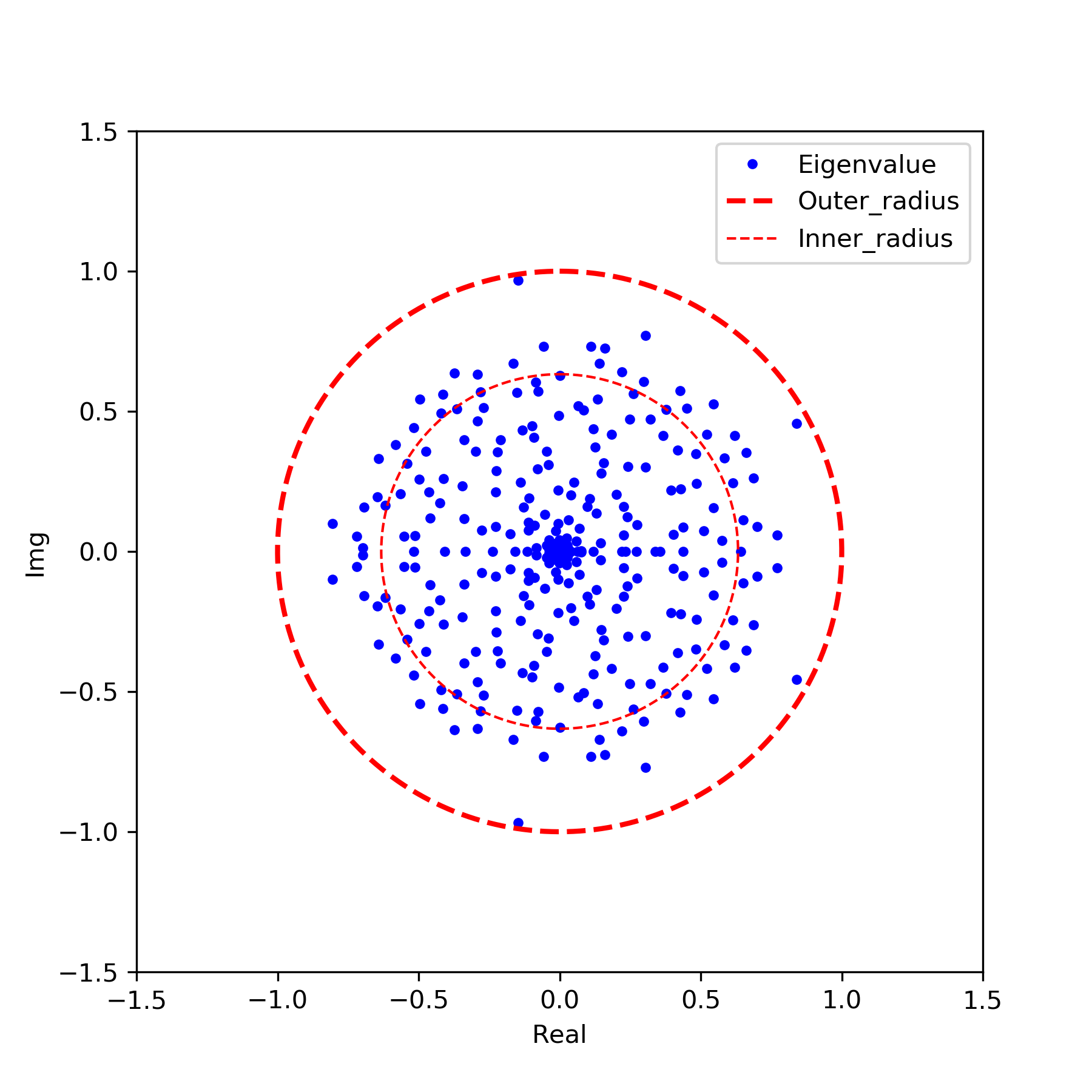}
}
\parbox{5cm}{\small \hspace{1.5cm}(b) $t_s=1001$}
\end{minipage}
\caption{The ESD and its comparison with theoretical Ring law in Case A.}
\label{fig:case1_law}
\end{figure}
\uppercase\expandafter{\romannumeral1}. During $t_s=500\sim 1000$, $\kappa_{\rm {MSR}}$ and $\eta$ remain almost constant and the corresponding values of $1-\alpha$ are small, which indicates no anomalies occur and the system operates in normal state. For example, at $t_s=1000$, the calculated value of $1-\alpha$ is $19.59\%$. As is shown in Figure \ref{fig:case1_law}(a), the ESD converges almost surely to the theoretical Ring law.

\uppercase\expandafter{\romannumeral2}. At $t_s=1001$, $\kappa_{\rm {MSR}}$ and $\eta$ begin to change rapidly and the corresponding $1-\alpha$ is $99.99\%$, which indicates an anomaly is detected and the system operates in abnormal state. As is shown in Figure \ref{fig:case1_law}(b), the ESD does not converge to the Ring law. It is noted that, from $t_s=1001\sim 1500$, the $\kappa_{\rm {MSR}}-t$ curve is almost U-shaped, because the delay lag of the anomaly signal to $\kappa_{\rm {MSR}}$ is equal to the moving window's width.

\uppercase\expandafter{\romannumeral3}. At $t_s=1501$, $\kappa_{\rm {MSR}}$ returns to normal and remains constant afterwards, which indicates the anomaly signal disappears and the system returns to normal state. It is noted that $\eta{(1501)}$ is large and the corresponding $1-\alpha$ is $99.99\%$, which is caused by the change of the system state.
\subsection{Case Study on $L>1$}
\label{subsection: case_B}
\begin{figure}[!t]
\centerline{
\includegraphics[width=3.0in]{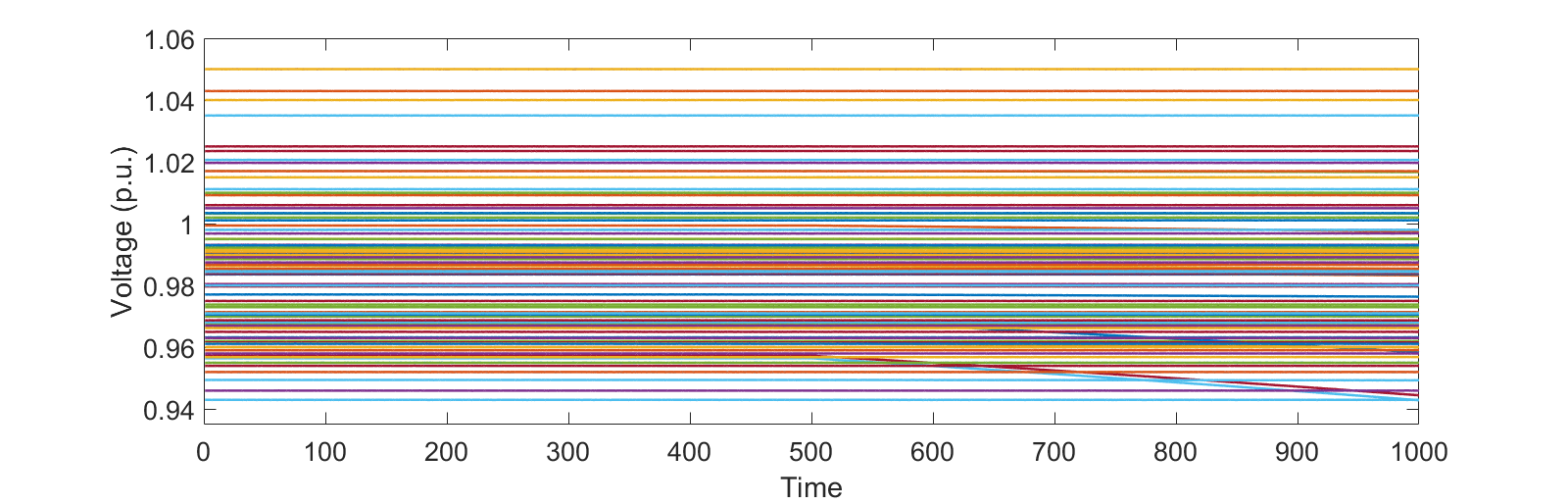}
}
\caption{The synthetic data generated from IEEE 118-bus test system in Case B. One increasing signal was set at $t_s=501$.}
\label{fig:case2_data_org}
\end{figure}

In this case, we test the effectiveness of our approach with $L>1$ and compare that with the case $L=1$. The synthetic data generated from IEEE 118-bus test system contained $118$ voltage measurement variables with sampling $1000$ times and a little white noise was introduced to represent random fluctuations and measuring errors. In the simulation, an increasing anomaly signal was set by a gradual increase of active load at bus $20$ and others stayed unchanged, which was shown in Table \ref{Tab: Case2}. The generated data was shown in Figure \ref{fig:case2_data_org}. In the experiments, the size of the moving window was set as $118\times 200$ and the signal-noise-rate $\tau_{SNR}$ was set to be $10000$.. The experiments were conducted for $20$ times and the results were averaged.
\begin{table}[!t]
\caption{An Assumed Signal for Active Load of Bus $20$ in Case B.}
\label{Tab: Case2}
\centering
\begin{tabular}{clc}   
\toprule[1.0pt]
\textbf {Bus} & \textbf{Sampling Time}& \textbf{Active Power(MW)}\\
\hline
\multirow{2}*{20} & $t_s=1\sim 500$ & $20$ \\
~&$t_s=501\sim 1000$ & $20\rightarrow 70$ \\
Others & $t_s=1\sim 1000$ & Unchanged \\
\hline
\end{tabular}
\end{table}

\begin{figure}[!t]
\centering
\begin{minipage}{4.1cm}
\centerline{
\includegraphics[width=1.8in]{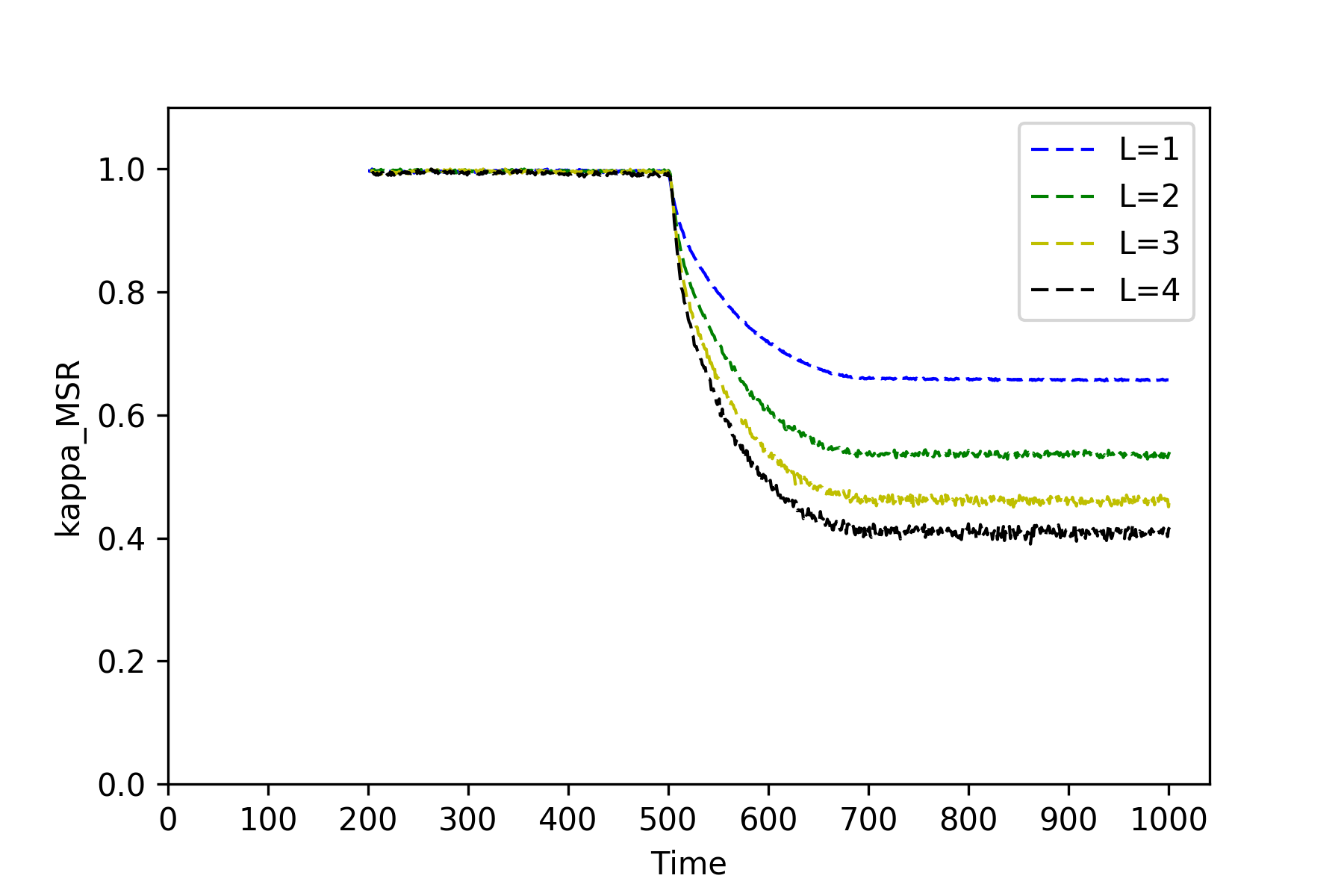}
}
\parbox{5cm}{\small \hspace{1.2cm}(a) $\kappa_{\rm MSR}-t$ curve}
\end{minipage}
\hspace{0.2cm}
\begin{minipage}{4.1cm}
\centerline{
\includegraphics[width=1.8in]{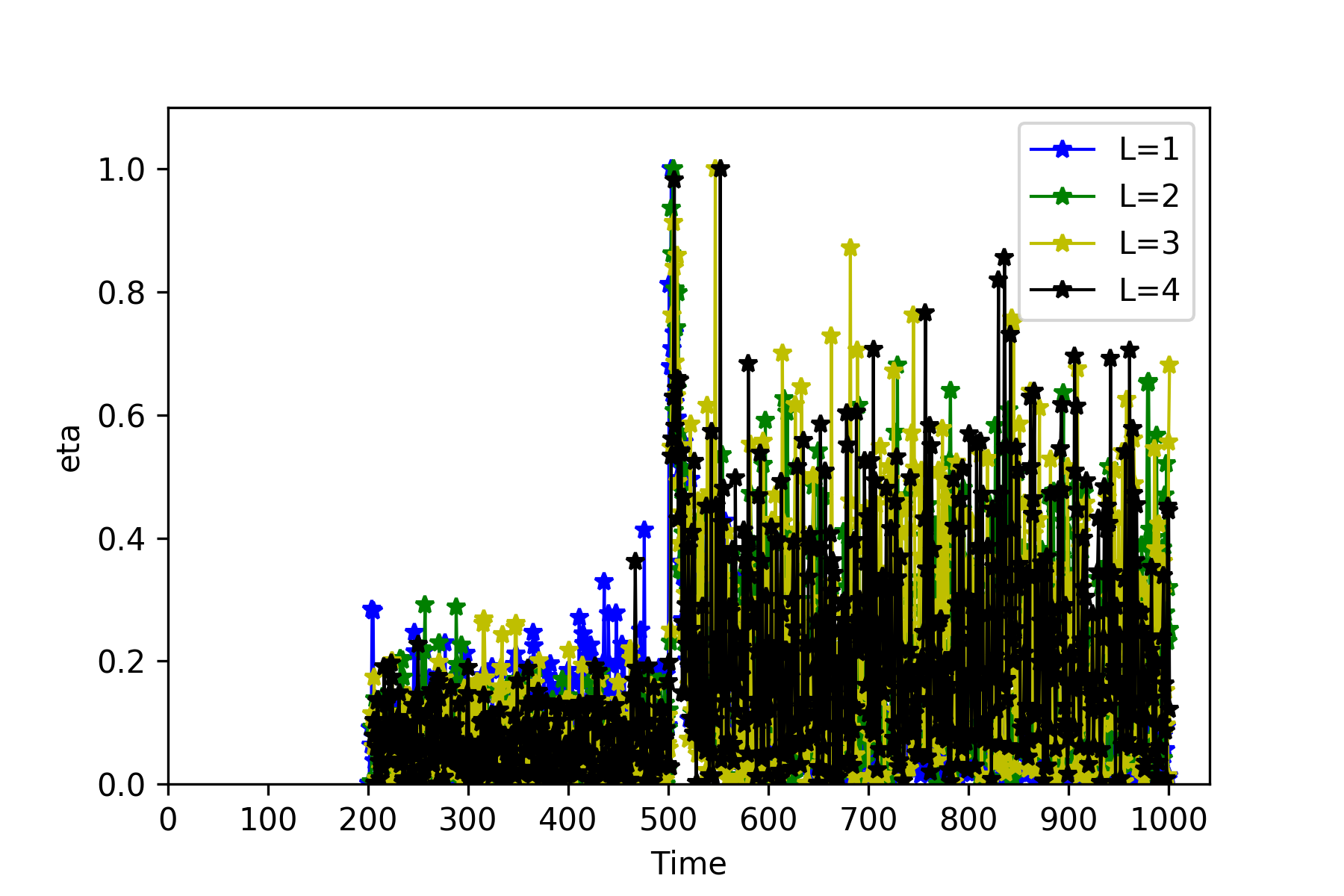}
}
\parbox{5cm}{\small \hspace{1.5cm}(b) $\eta-t$ curve}
\end{minipage}
\caption{The anomaly detection results in Case B.}
\label{fig:case2_indicator}
\end{figure}
The anomaly detection results of our approach with $L=1\sim 4$ are shown in Figure \ref{fig:case2_indicator}, where $\kappa_{\rm MSR}$ and $\eta$ are normalized into $(0,1]$. It is noted that the $\kappa_{\rm {MSR}}-t$ curves corresponding to $L=1,2,3,4$ begin at $t_s=200,201,202,203$, because the initial moving window contains $200$ times of sampling data and the initial $\kappa_{\rm {MSR}}$ is calculated through the product of $L$ consecutive moving data windows. And the $\eta-t$ curves corresponding to $L=1,2,3,4$ begin at $t_s=201,202,203,204$, because the initial $\eta{(t_k)}$ is calculated through $\kappa _{\rm {MSR}}{(t_{k-1})}$ and $\kappa _{\rm {MSR}}{(t_k)}$ in equation (\ref{Eq:eta}). In real-time analysis, the parameter $T$ and the threshold ${(1-\alpha)}_{th}$ are set the same as in Case A. The anomaly detection processes are shown as follows:

\begin{figure*}[!t]
\centering
\begin{minipage}{4.1cm}
\centerline{
\includegraphics[width=1.8in]{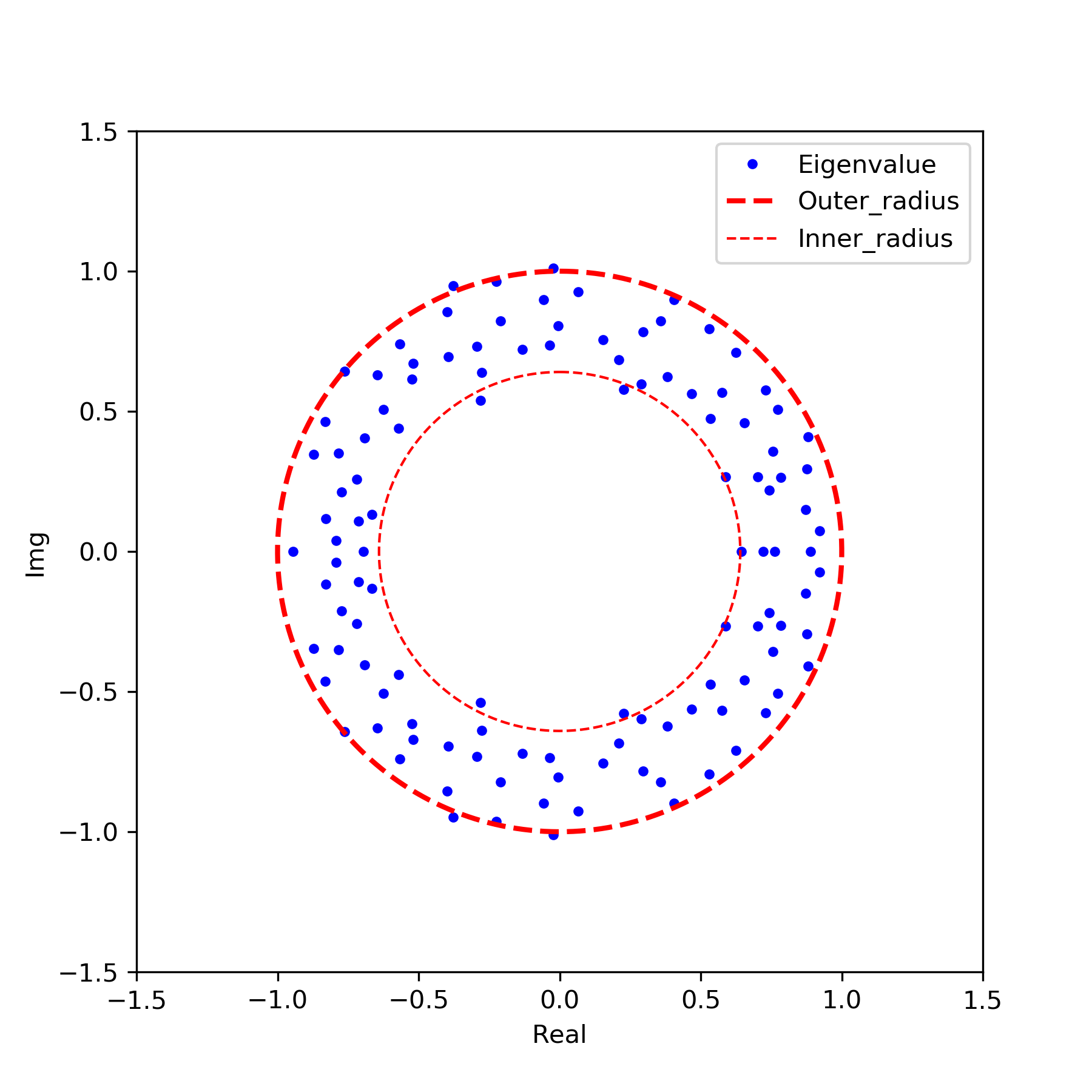}
}
\parbox{5cm}{\small \hspace{1.5cm}(a) $L=1$}
\end{minipage}
\hspace{0.2cm}
\begin{minipage}{4.1cm}
\centerline{
\includegraphics[width=1.8in]{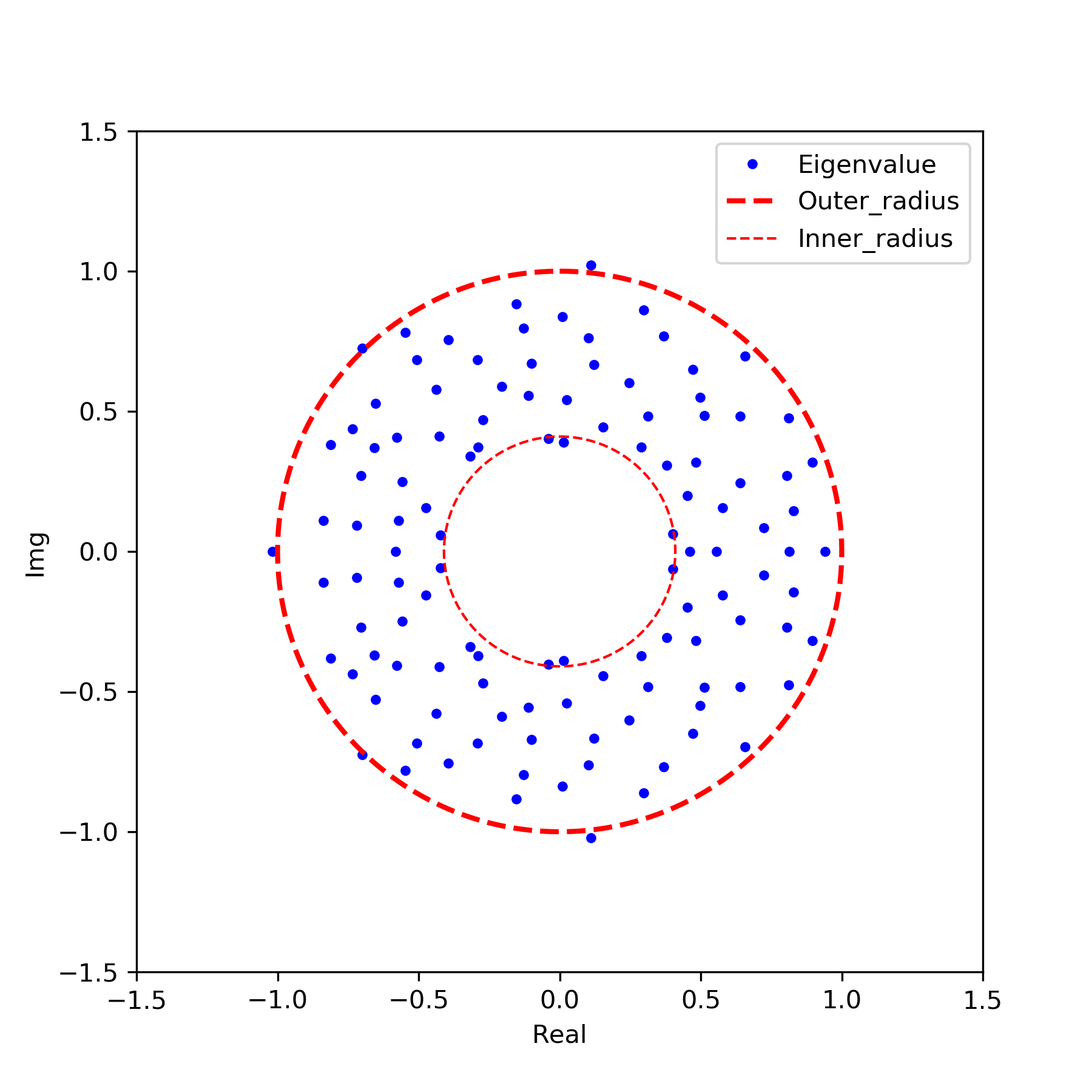}
}
\parbox{5cm}{\small \hspace{1.5cm}(b) $L=2$}
\end{minipage}
\hspace{0.2cm}
\begin{minipage}{4.1cm}
\centerline{
\includegraphics[width=1.8in]{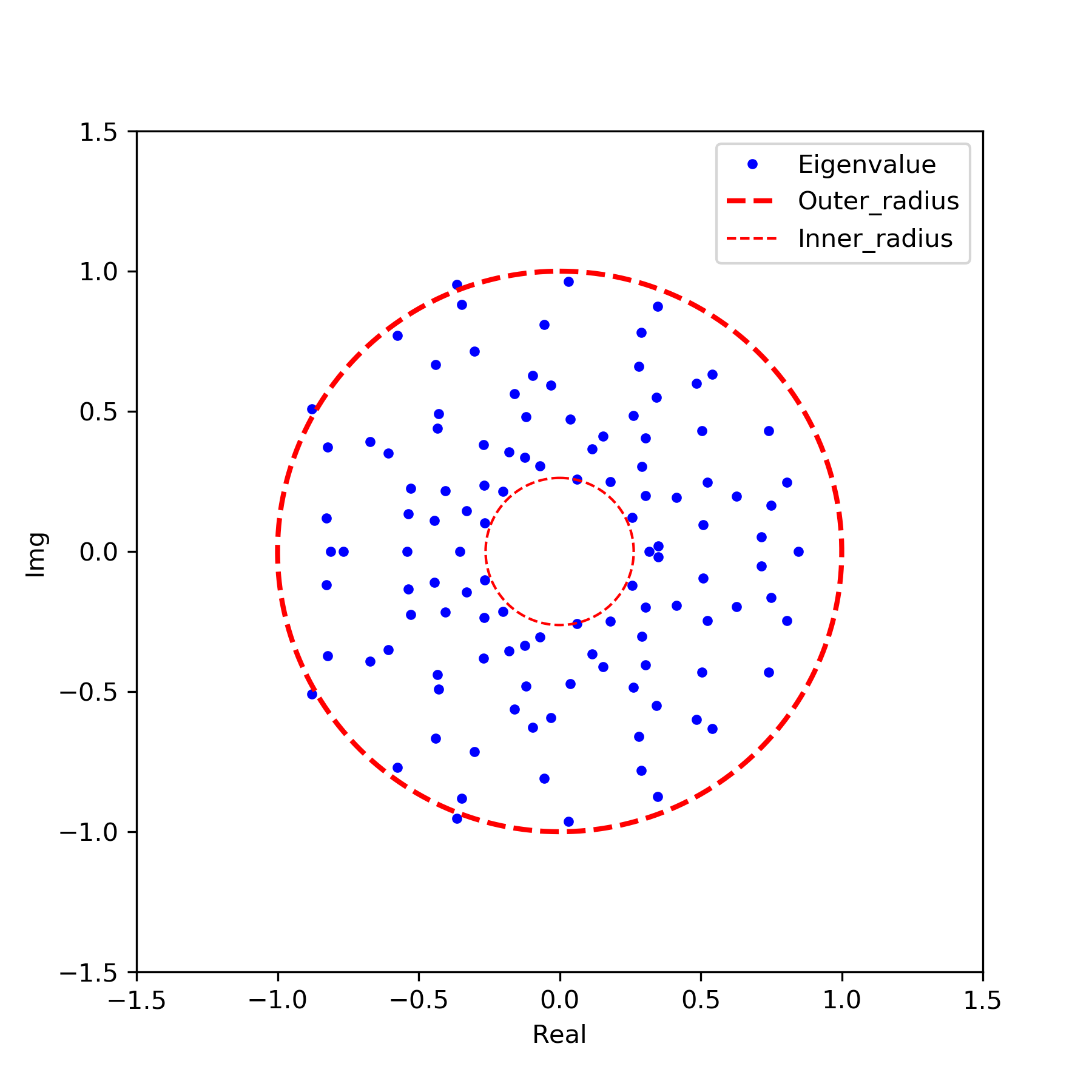}
}
\parbox{5cm}{\small \hspace{1.5cm}(c) $L=3$}
\end{minipage}
\hspace{0.2cm}
\begin{minipage}{4.1cm}
\centerline{
\includegraphics[width=1.8in]{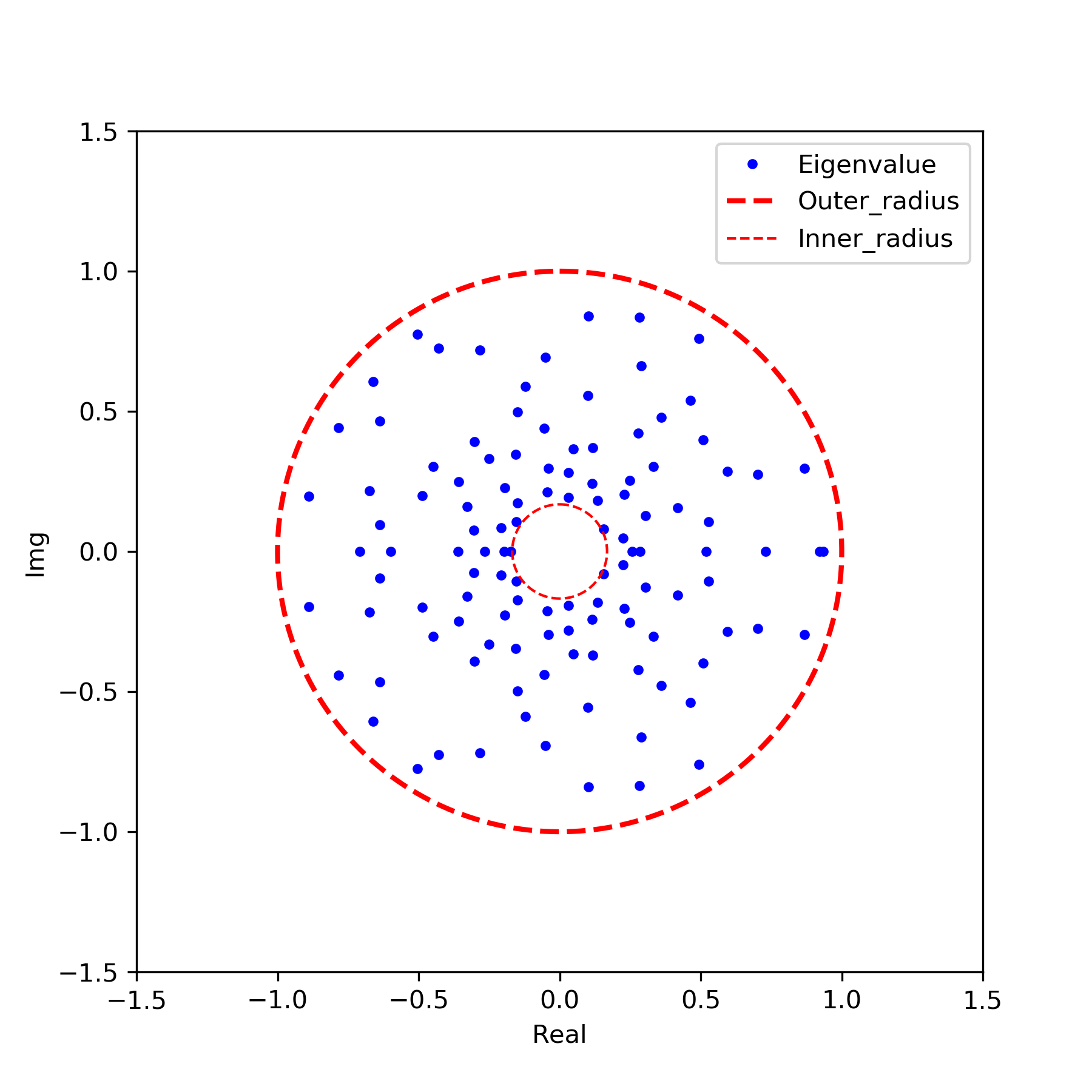}
}
\parbox{5cm}{\small \hspace{1.5cm}(d) $L=4$}
\end{minipage}
\caption{In normal state (such as $t_s=500$), the ESD converges almost surely to the theoretical Ring law.}
\label{fig:case2_law_normal}
\end{figure*}
\uppercase\expandafter{\romannumeral1}. During $t_s=200\sim 500$, $\kappa_{\rm {MSR}}$ and $\eta$ remain almost constant and the corresponding $1-\alpha$ of $\eta$ are small random values, which indicates no anomalies occur and the system operates in normal state. For example, at $t_s=500$, the calculated $1-\alpha$ of $\eta$ corresponding to $L=1,2,3,4$ are $45.457\%, 45.045\%, 39.943\%, 12.609\%$, respectively. As is shown in Figure \ref{fig:case2_law_normal}, the ESD converges almost surely to the theoretical Ring law.

\begin{figure*}[!t]
\centering
\begin{minipage}{4.1cm}
\centerline{
\includegraphics[width=1.8in]{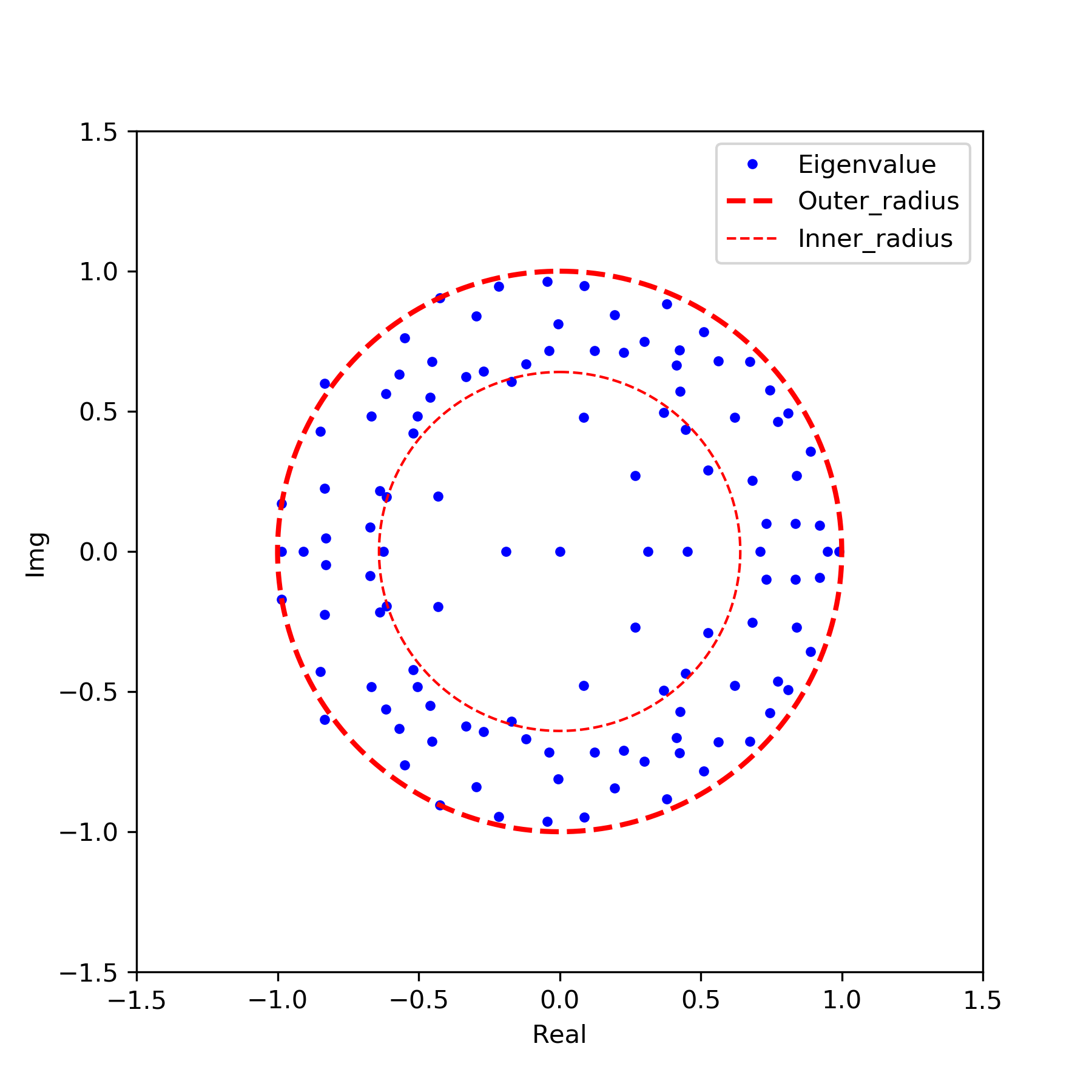}
}
\parbox{5cm}{\small \hspace{1.5cm}(a) $L=1$}
\end{minipage}
\hspace{0.2cm}
\begin{minipage}{4.1cm}
\centerline{
\includegraphics[width=1.8in]{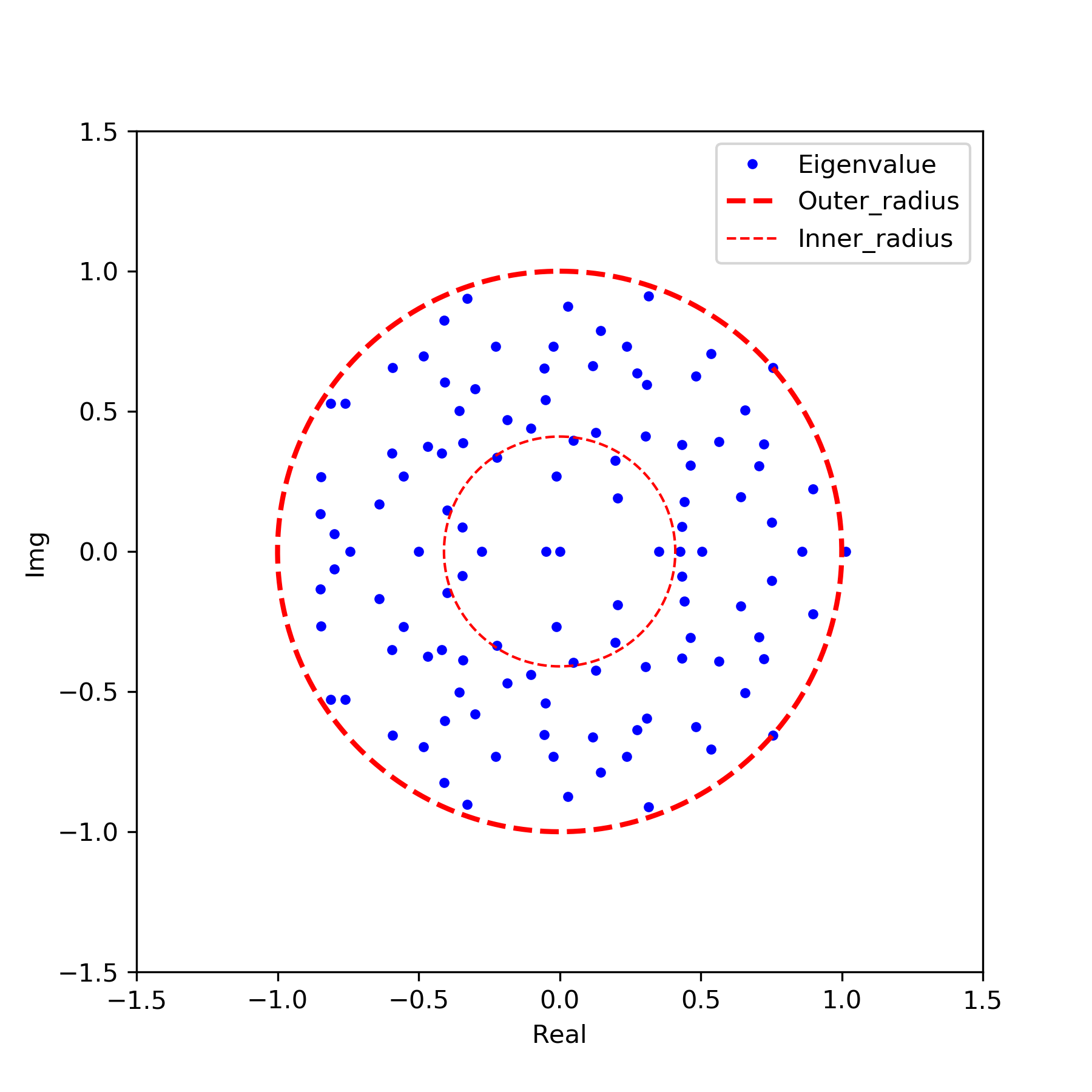}
}
\parbox{5cm}{\small \hspace{1.5cm}(b) $L=2$}
\end{minipage}
\hspace{0.2cm}
\begin{minipage}{4.1cm}
\centerline{
\includegraphics[width=1.8in]{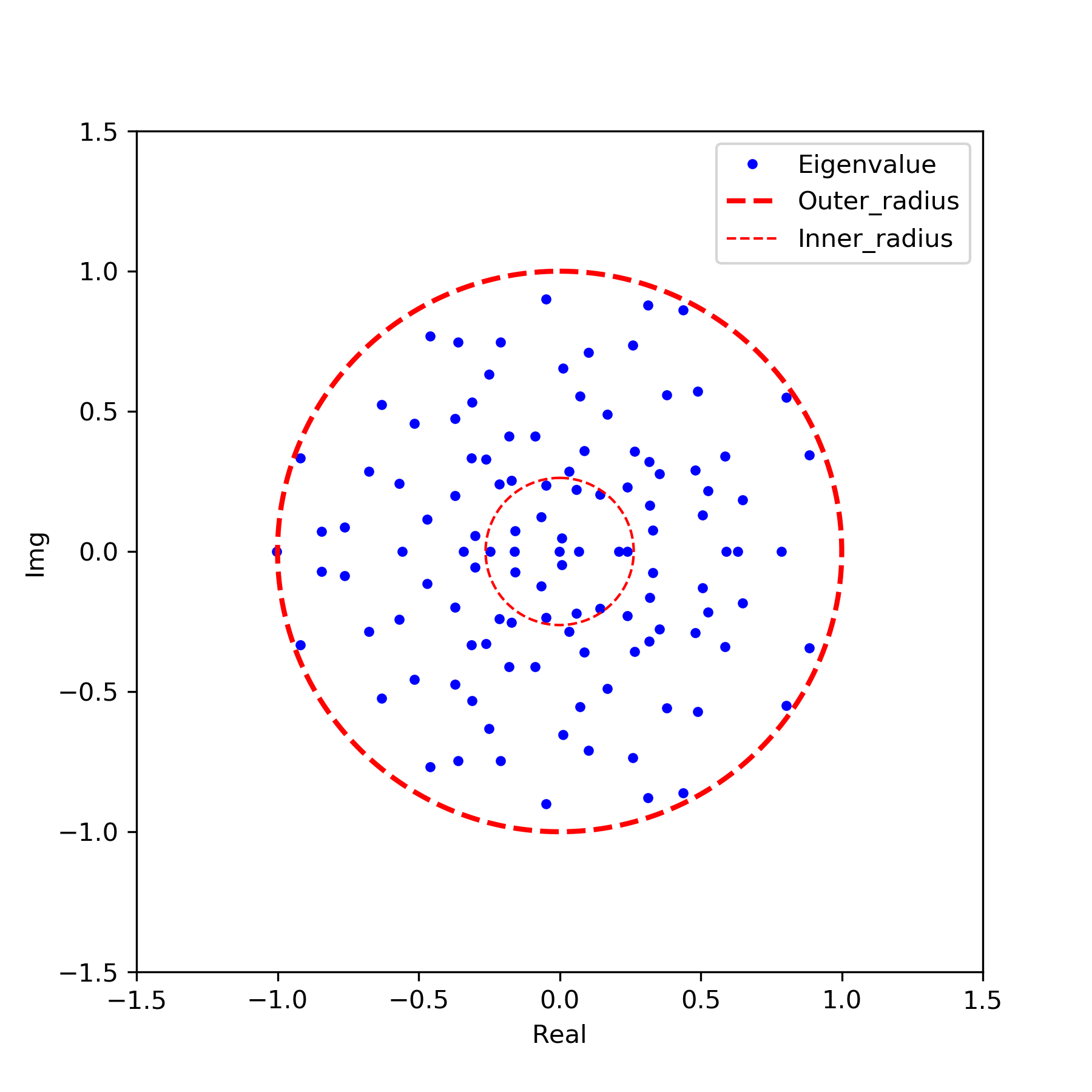}
}
\parbox{5cm}{\small \hspace{1.5cm}(c) $L=3$}
\end{minipage}
\hspace{0.2cm}
\begin{minipage}{4.1cm}
\centerline{
\includegraphics[width=1.8in]{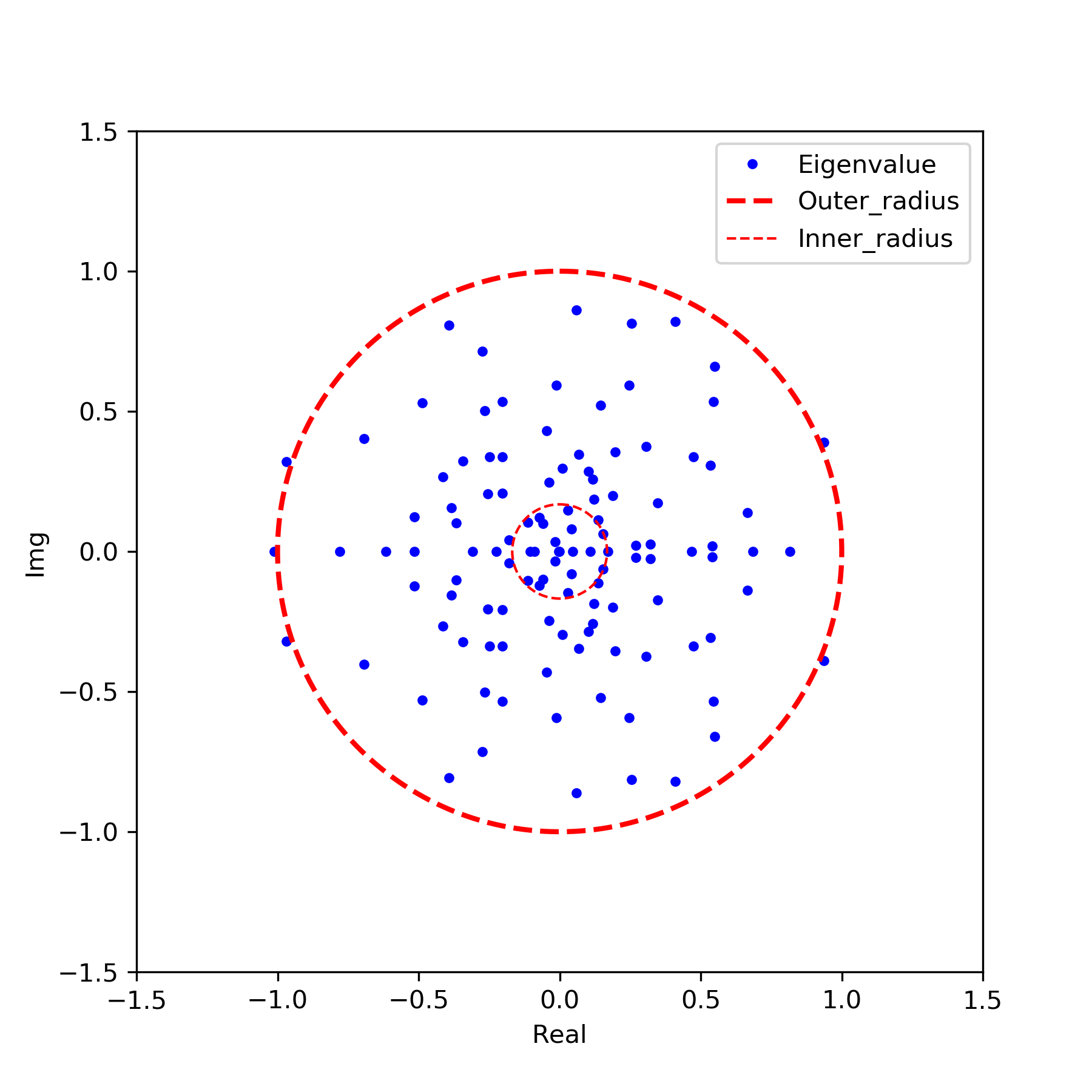}
}
\parbox{5cm}{\small \hspace{1.5cm}(d) $L=4$}
\end{minipage}
\caption{In abnormal state (such as $t_s=501$), the ESD does not converge to the theoretical Ring law.}
\label{fig:case2_law_abnormal}
\end{figure*}
\uppercase\expandafter{\romannumeral2}. At $t_s=501$, $\kappa_{\rm {MSR}}$ and $\eta$ change rapidly and the corresponding $1-\alpha$ of $\eta$ for $L=1,2,3,4$ are $99.998\%, 99.998\%, 99.999\%, 99.999\%$, which indicates an anomaly is detected and the system operates in abnormal state. As is shown in Figure \ref{fig:case2_law_abnormal}, the ESD does not converge to the Ring law. It is noted that, with the increase of the number of multiplicative data windows, the values of $\kappa_{\rm {MSR}}$ decrease more rapidly from $t_s=501\sim 700$, which makes it easier for detecting the anomaly signal. It validates our claim in Section \ref{subsection: statistical properties} that the product of multiple `signal+noise' matrices can reinforce the signals.

\uppercase\expandafter{\romannumeral3}. From $t_s=701$, $\kappa_{\rm {MSR}}$ and $\eta$ remain almost constant for the increasing signal is contained throughout the moving window afterwards.
\subsection{Case Study on Comparison with Other Approaches}
\label{subsection: case_C}
\begin{figure}[!t]
\centerline{
\includegraphics[width=2.5in]{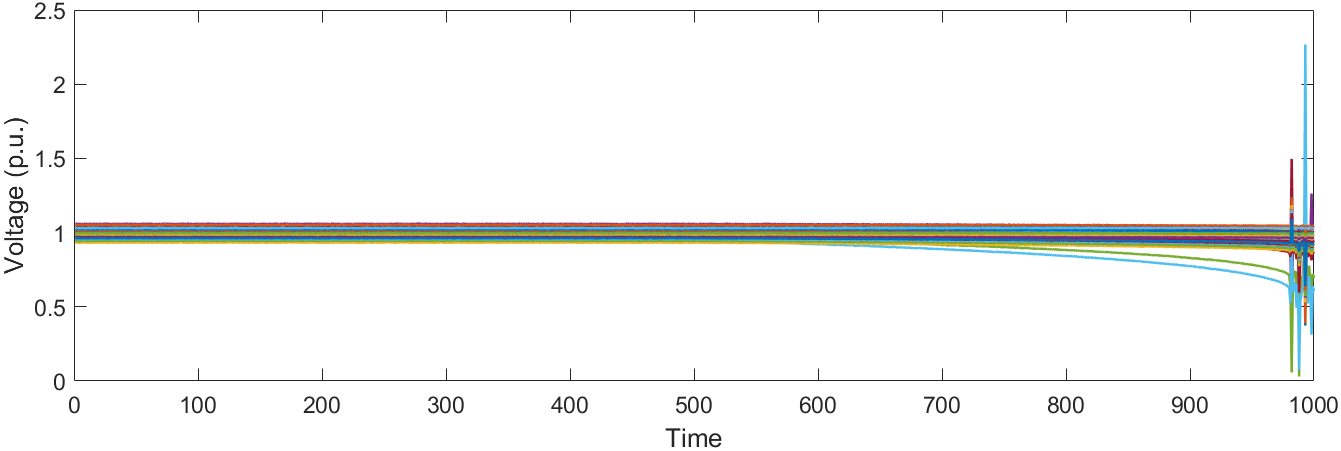}
}
\caption{The synthetic data generated from IEEE 57-bus test system. An increasing anomaly signal was set at $t_s=501$. With the increase of the anomaly signal, the voltage collapses at $t_s=980$.}
\label{fig:case3_data_org}
\end{figure}

In this case, we compare our spectrum analysis (SA) approach with one-class support vector machines (SVMs) \cite{ma2003time}, structured autoencoders (AEs) \cite{liu2018anomaly} and long short term memory (LSTM) networks \cite{malhotra2015long} to illustrate the advantages of our approach for anomaly detection, i.e., more sensitive to the variation of the data behavior and robust against random fluctuations and measuring errors. The IEEE 57-bus test system is used to generate the synthetic data. In the simulation, an increasing anomaly signal was set by a gradual increase of active load at bus $20$ and others stayed unchanged, which was shown in Table \ref{Tab: Case3}. The generated data contained $57$ voltage measurement variables with sampling $1000$ times, which was shown in Figure \ref{fig:case3_data_org}. In the experiment, the signal-noise-rate $\tau_{SNR}$ was set as $1000$. For SVMs, AEs and LSTM, we trained the detection models only using a normal data sequence during $t_s=1\sim 200$ and computed the testing errors for the remaining sequence during $t_s=201\sim 1000$, in which one times of sampling was used as a training/testing sample. For our approach, both the case of $L=1$ and $L>1$  were tested. The parameters involved in the detection approaches are summarized as in Table \ref{Tab: Case3_parameter}.
\begin{table}[!t]
\caption{An Assumed Signal for Active Load of Bus $20$ in Case C.}
\label{Tab: Case3}
\centering
\begin{tabular}{clc}   
\toprule[1.0pt]
\textbf {Bus} & \textbf{Sampling Time}& \textbf{Active Power(MW)}\\
\hline
\multirow{2}*{20} & $t_s=1\sim 500$ & $10$ \\
~&$t_s=501\sim 1000$ & $10\rightarrow 60$ \\
Others & $t_s=1\sim 1000$ & Unchanged \\
\hline
\end{tabular}
\end{table}
\begin{table}[!t]
\caption{Parameter Settings Involved in the Detection Approaches.}
\label{Tab: Case3_parameter}
\centering
\footnotesize
\begin{tabular}{p{1.8cm}p{5.8cm}}   
\toprule[1.0pt]
\textbf{Approaches} & \textbf{Parameter Settings} \\
\hline
\multirow{2}*{SVMs} & the upper bound on the fraction of training errors $v$: 0.03; \\
~&the kernel function: $K({{\bf x}_i},{{\bf x}_j})={(0.01{{\bf x}_i}^{T}{{\bf x}_j})}^{3}$; \\
\hline
\multirow{7}*{AEs} & the model depth: $3$; \\
~&the number of neurons in each layer of encoder: $57,32,16$; \\
~&the number of neurons in each layer of decoder: $16,32,57$; \\
~&the initial learning rate: $0.001$;  \\
~&the activation function: $sigmoid$; \\
~&the minimum reconstruction error: $0.00001$;  \\
~&the optimizer: $Adam$.  \\
\hline
\multirow{7}*{LSTM} & the time steps: $1$; \\
~&the model depth: $3$; \\
~&the number of neurons in each layer: $57,64,57$; \\
~&the initial learning rate: $0.001$;  \\
~&the activation function: $sigmoid, tanh$; \\
~&the minimum loss: $0.00001$;  \\
~&the optimizer: $Adam$.  \\
\hline
\multirow{2}*{SA($L=1$)} & the moving window's size: $57\times 200$; \\
~&the number of multiplicative data windows: $L=1$. \\
\hline
\multirow{2}*{SA($L>1$)} & the moving window's size: $57\times 200$; \\
~&the number of multiplicative data windows: $L=2$. \\
\hline
\end{tabular}
\end{table}

\begin{figure}[!t]
\centerline{
\includegraphics[width=3.0in]{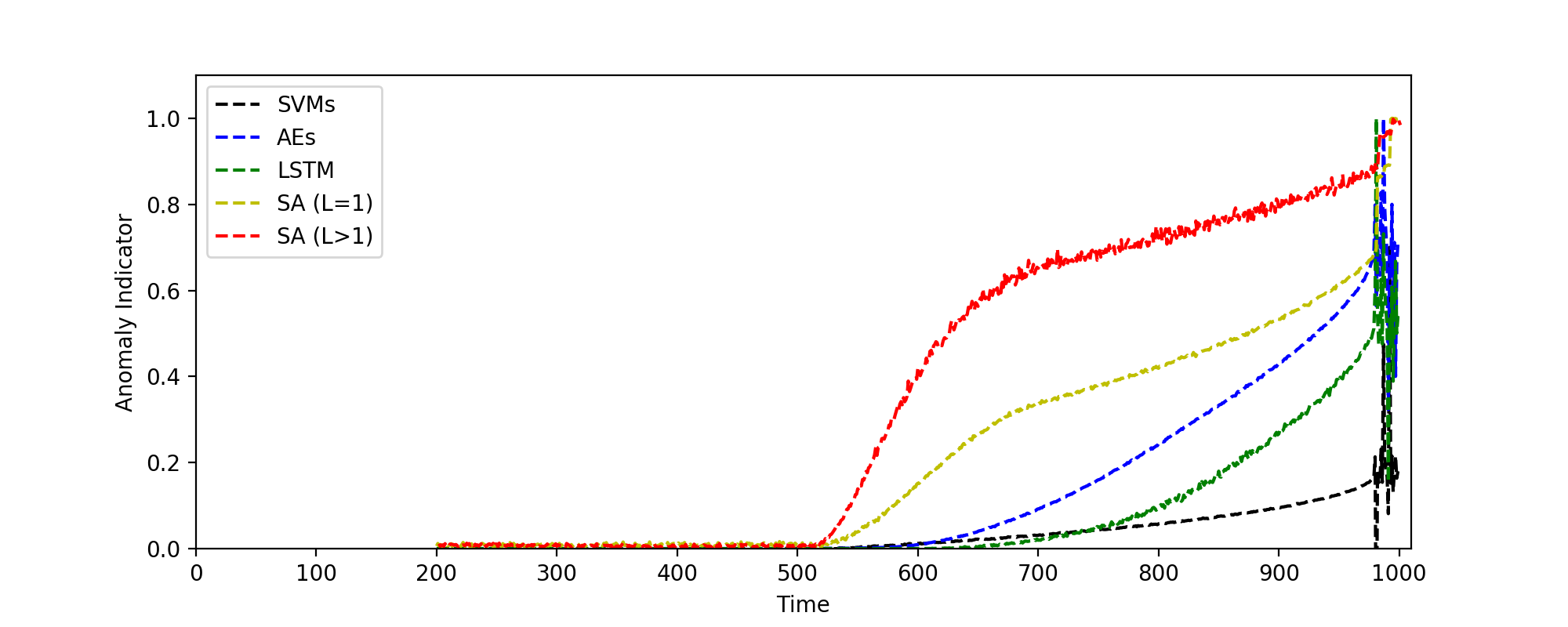}
}
\caption{The anomaly detection results of different approaches in Case C.}
\label{fig:case3_comparison}
\end{figure}
The anomaly detection results of different approaches are normalized into $[0,1]$, which are shown in Figure \ref{fig:case3_comparison}. For SVMs, the normalization result of signed distance to the separating hyperplane is plotted;  for AEs and LSTM, the normalized values of testing errors are plotted; for SA ($L=1$) and SA ($L>1$), $1-{\hat\kappa}_{\rm {MSR}}$ is plotted, where ${\hat\kappa}_{\rm {MSR}}$ is the normalized value of ${\kappa}_{\rm {MSR}}$. It can be obtained that:

\uppercase\expandafter{\romannumeral1}. SA ($L=1$) and SA ($L>1$) are able to detect the anomaly signal much earlier (i.e., $t_s=520\sim 530$) than other approaches, which validates our approach is more sensitive to the variation of the data behavior by exploring the data correlations and robust against random fluctuations and measuring errors. The reason is that, in our approach, a data window instead of just the current sampling data is analyzed for each sampling time. The average result makes our approach more robust against random fluctuations and measuring errors of the data.

\uppercase\expandafter{\romannumeral2}. SA ($L>1$) outperforms SA ($L=1$) in anomaly detection, which indicates the product of multiple moving data windows can reinforce the anomaly signal so that it can be detected much easier.

\uppercase\expandafter{\romannumeral3}. The detection curves in our SA approach increase rapidly when the voltage collapses from $t_s=980$, while that of other approaches vibrates with the voltage values. It indicates our approach can reflect the system state more accurately in macroscopic.

\begin{table}[!t]
\caption{The Average Calculation Time of Different Approaches.}
\label{Tab: Case3_comparison}
\centering
\begin{tabular}{cccccc}   
\toprule[1.0pt]
\textbf{Approaches} & {SVMs} & {AEs} & {LSTM} & {SA($L=1$)} & {SA($L=2$)} \\
\hline
{$ACT$ (unit: s)} & {$0.001$} & {$0.001$} & {$0.002$} & {$0.001$} & {$0.002$} \\
\hline
\end{tabular}
\end{table}
Furthermore, in order to illustrate the efficiency of different detection approaches, the $average\; calculation\; time \; (ACT)$ for each sampling time was counted. For SVMs, AEs and LSTM, the $ACT$ for each testing sample was counted, which does not include the model training time. The experiments were conducted on a server with $2.60$ GHz central processing unit (CPU) and $8.00$ GB random access memory (RAM). The $ACT$ for SVMs, AEs, LSTM, SA ($L=1$) and SA ($L=2$) are shown in Table \ref{Tab: Case3_comparison}. Considering our approach is an unsupervised approach without any training, it can be concluded that our approach has competitive performance in detection efficiency.

\section{Conclusion}
\label{section: conclusion}
Based on random matrix theory, a data-driven approach is proposed for anomaly detection in power systems. It is capable of detecting  the anomaly in an early phase by exploring the variation of the data behavior. The mean spectral radius gives insight into the system states from a macroscopic perspective, which is used to indicate the data behavior in our approach. An anomaly indicator based on the MSR is designed and the corresponding confidence level $1-\alpha$ is calculated to realize anomaly declare automatically. Our approach is purely data-driven without making assumptions and simplifications on the complex power systems. It is robust against random fluctuations and measuring errors, and it has fast computing speed. Case studies on synthetic data corroborate the effectiveness and advantages of our approach, which indicates our approach can be served as a primitive for real-time data analysis in power systems.

\small{}
\bibliographystyle{IEEEtran}
\bibliography{helx}

\normalsize{}
\end{document}